\documentclass[a4paper,12pt,preprint,nofootinbib]{revtex4}

\usepackage{graphicx}
\usepackage{amsmath}
\usepackage{amssymb}
\usepackage{amsthm}
\usepackage{tikz}
\usepackage{mathrsfs}
\usepackage{multirow}
\usepackage{scalerel}
\usepackage{mathtools}
\usepackage{textcomp}
\usepackage{dsfont}
\usepackage{blkarray}
\usepackage{bm}
\usepackage{todonotes}
\usepackage{hyperref}
\usepackage{pgf-pie}
\usepackage[compat=1.1.0]{tikz-feynman}
\allowdisplaybreaks[4]
\definecolor{darkyellow}{rgb}{0.5, 0.5, 0.0}
\definecolor{darkpurple}{rgb}{0.5, 0.2, 0.8}
\definecolor{darkblue}{rgb}{0.0, 0.0, 0.8}
\definecolor{darkgreen}{rgb}{0.0, 0.4, 0.0}
\definecolor{darkred}{rgb}{0.5, 0.0, 0.0}
\hypersetup{
    linktocpage,
     colorlinks,
     citecolor=darkgreen,
     linkcolor= darkgreen,
     urlcolor=darkgreen
}

\definecolor{col1}{rgb}{0.81, 0.85, 0.91}
\definecolor{col2}{rgb}{0.96, 0.88, 0.74}
\definecolor{col3}{rgb}{0.87, 0.91, 0.76}
\definecolor{col4}{rgb}{0.98, 0.82, 0.76}
\definecolor{col5}{rgb}{0.86, 0.84, 0.91}
\definecolor{col6}{rgb}{0.93, 0.83, 0.73}

\newcommand{\LQCD}{{\Lambda_\text{QCD}}}
\newcommand{\onebar}{{\bar{1}}}
\newcommand{\twobar}{{\bar{2}}}
\newcommand{\threebar}{{\bar{3}}}
\newcommand{\ibar}{{\bar{i}}}

\newcommand{\cusp}{{\text{cusp}}}
\newcommand{\etal}{{{\eta_{\ell}}}}
\newcommand{\etah}{{{\eta_{h}}}}

\newcommand{\sub}{ {\text{sub}}}

\newcommand{\cO}{{\mathcal O}}

\newcommand{\cM}{{\mathcal M}}

\newcommand{\hemi}{\text{hemi}}
\newcommand{\alphs}[1]{\frac{\alpha_s(#1)}{4\pi}}
\tikzset{
    position/.style args={#1:#2 from #3}{
        at=(#3.#1), anchor=#1+180, shift=(#1:#2)
    }
}

\begin{document}

\title{\center Sudakov Shoulder Resummation for 
\\
Thrust and Heavy Jet Mass} 

\author{Arindam Bhattacharya}
\email{arindamb@g.harvard.edu}
\author{Matthew~D.~Schwartz}
\email{schwartz@g.harvard.edu}
\author{Xiaoyuan Zhang}
\email{xiaoyuanzhang@g.harvard.edu}

\affiliation{Department of Physics, Harvard University, Cambridge, MA 02138, USA}

\begin{abstract}
When the allowed range of an observable grows order-by-order in perturbation theory, its perturbative expansion can have discontinuities (as in the $C$ parameter) or discontinuities in its derivatives (as in thrust or heavy jet mass) called Sudakov shoulders. 
 We explore the origin of these logarithms using both perturbation theory and effective field theory.
We show that for thrust and heavy jet mass, the logarithms arise from kinematic configurations with narrow jets and deduce the next-to-leading logarithmic series. 
The left-shoulder logarithms in heavy jet mass ($\rho)$ of the form $r\ \alpha_s^n \ln^{2n}r $ with $r=\frac{1}{3}-\rho$ are particularly dangerous, because they invalidate fixed order perturbation theory in regions traditionally used to extract $\alpha_s$. Although the factorization formula shows there are no non-global logarithms, we find Landau-pole like singularities in the resummed distribution associated with the cusp anomalous dimension, and that power corrections are exceptionally important. 

\end{abstract}

\maketitle

\section{Introduction}
\begin{figure}[t]
    \centering
    \includegraphics[scale=1]{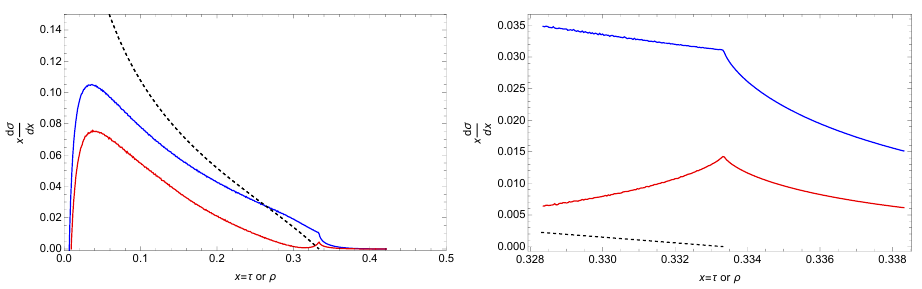}
    \caption{Thrust (blue) and heavy jet mass (red) at NLO compared to LO (dashed). The NLO curve does not have LO added in. That is, the LO is the $\frac{\alpha_s}{2\pi}$ times the ``A'' function and the NLO curves are $(\frac{\alpha_s}{2\pi})^2$ times the ``B'' functions, in the notation of~\cite{Gehrmann-DeRidder:2007vsv}.
    Right is a zoom-in of Sudakov shoulder region near  $\frac{1}{3}$. The NLO computation is performed with the program {\sc event2} \cite{Catani:1996jh,Catani:1996vz}. All distributions are normalized to Born cross section $\sigma_0$.
    \label{fig:thrustandhjm}}
\end{figure}

It is not uncommon for an observable to have a range that grows order-by-order in perturbation theory. Traditional $e^+e^-$ event shapes, such as thrust, the $C$ parameter, and heavy jet mass~\cite{Catani:1997xc} have this property as do some hadron-collider observables like the jet shape~\cite{Seymour:1997kj,Luisoni:2020efy}. Similar behavior can also be seen in the soft-drop jet mass~\cite{Benkendorfer:2021unv}.
As observed by Catani and Webber~\cite{Catani:1997xc}, when the range grows order-by-order,
there can be incomplete cancellations between the virtual contributions, which are confined to the lower-order range, and the real-emission contributions, which are not. The results are distributions with non-analytic behavior at intermediate values of the observable: discontinuities, cusps or kinks at any given finite order in perturbation theory, collective called {\it Sudakov shoulders}, as shown in Fig.~\ref{fig:thrustandhjm}.  Sudakov shoulders are caused by large logarithms associated with kinematic regions not close to the absolute (non-perturbative) phase space boundary. We classify the Sudakov shoulders as either right shoulders, which have large logarithms extending into regions accessible only at higher orders in perturbation theory (i.e. to the right of the shoulder as in thrust or $C$ parameter) or left shoulders, which have logarithms affecting regions accessible at all orders in perturbation theory (i.e. to the left of the shoulder, as in heavy jet mass).
Left shoulders are particularly problematic as they can invalidate the use of fixed order perturbation theory over a wide range of observable values. 

\begin{figure}[t]
    \centering
    \includegraphics[scale=1]{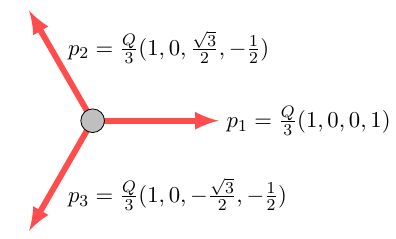}
    \caption{The trijet configuration where $\rho=\tau=\frac{1}{3}$ has 3 equally spaced jets of equal energy}
    \label{fig:trijet}
\end{figure}

To understand Sudakov shoulders, consider first the thrust observable~\cite{Farhi:1977sg}. Thrust is defined in the center-of-mass frame of an $e^+ e^-$ collision as
\begin{equation}
  T \equiv \max_{\vec{n}} \frac{\sum_j | \vec{p}_j \cdot \vec{n} |}{\sum_j |
  \vec{p}_j |}
  \label{tdef}
\end{equation}
where the sum is over all particles in the event and the maximum is over
3-vectors $\vec{n}$ of unit norm. It is common to use $\tau = 1 - T$ in place of $T$. 
 The vector $\vec{n}$ that maximizes thrust is
known as the {\it thrust axis}.
When there are only 2 particles, they must be back-to-back, and then $\tau = 0$ exactly. If there are 3 massless particles, then the phase space is 2 dimensional and can be parameterized with $s_{ij} = (p_i + p_j)^2/Q^2$ constrained by $s_{12} + s_{23} + s_{13}=1$ with $Q$ the center-of-mass energy. Then
\begin{equation}
    \tau= \min (s_{12}, s_{13}, s_{23}) \le \frac{1}{3}
\end{equation}
The phase space point that saturates this bound has $s_{12}=s_{13}=s_{23}=\frac{1}{3}$ and comprises the symmetric {\it trijet} configuration: 3 particles of equal energy and angular separation, as shown in Fig~\ref{fig:trijet}. Near this point the spin-summed 3-body matrix-element-squared is not exceptional
\begin{equation}
| \cM_1 |^2= | \cM_0 |^2 2 C_F g_s^2 \frac{ s_{12}^2 + s_{23}^2 + 2Q^2 s_{13}}{s_{12} s_{13}}
\cong | \cM_0 |^2 64 \pi C_F \alpha_s
\label{M1form}
\end{equation}
where $|\mathcal{M}_0|^2$ is the $\gamma^{*}\to q \bar q$ matrix-element-squared. Because the phase space goes to zero at $\tau=\frac{1}{3}$, the differential cross section must vanish there. The result is that
\begin{equation}
    \frac{1}{\sigma_0} \frac{d\sigma}{d\tau} \cong  3\times 48 C_F\frac{\alpha_s}{4\pi} \int_0^{\frac{1}{3}-\tau} d s_{12}
    = 144 C_F \frac{\alpha_s}{4\pi}\left(\frac{1}{3} - \tau\right) \theta\left(\frac{1}{3} - \tau\right)
    \label{lothrust}
\end{equation}
with the factor of 3 coming from the 3 choices of thrust axis all of which contribute equally near $\tau=\frac{1}{3}$. Already here we can see the Sudakov shoulder: there is a discontinuity in the first derivative of the distribution from $-144 C_F \frac{\alpha_s}{4\pi}$ for $\tau<\frac{1}{3}$ to $0$ for $\tau>\frac{1}{3}$.

Given the thrust axis from the maximization in Eq.~\eqref{tdef}, the event is divided into two hemispheres. We can compute the invariant masses $m_1$ and $m_2$ of all the partons in hemisphere 1 and 2 and then heavy jet mass is defined as
\begin{equation}
    \rho = \frac{1}{Q^2} \max(m_1^2, m_2^2)
\end{equation}
At order $\alpha_s$ one hemisphere must be massless and $\tau=\rho$, and thus $\frac{d\sigma}{d\rho}$ has a discontinuity in its first derivative at leading order, just like $\tau$.

Now, consider what happens at higher order in perturbation theory. The parton in the light hemisphere will radiate gluons, making the light hemisphere massive. Since the cross section for the light jet having mass less than $m$ after one emission scales like $\sigma \sim \alpha_s \ln^2 m^2$ there is a Sudakov enhancement to the cross section at small $m^2$. As the light hemisphere jet grows, energy must be drawn away from the heavy hemisphere, making it lighter. Roughly speaking, setting $Q=1$ for simplicity, $\rho \lesssim \frac{1}{3} -m^2 $ (as we will derive). As a consequence, the cross section at $\rho = \frac{1}{3} -m^2 $ will be enhanced by factors of $\ln^2 m^2 = \ln^2(\frac{1}{3} - \rho)$.
Thus large Sudakov logs associated with radiation into the light hemisphere translate into Sudakov shoulder logs. This is the physical mechanism for the production of large logs in the left shoulder for heavy jet mass. 

To properly and systematically resum the Sudakov shoulder logarithms, we must understand this mechanism, as well as the consequences of radiation from the heavy-hemisphere partons. 
At first glance, the mechanism, which transfers large logs from the light to the heavy hemisphere using energy conservation may seem difficult to reconcile with factorization. 
Indeed, previous work has noted the recoil sensitivity of Sudakov shoulder logarithms starting at the next-to-leading logarithmic level \cite{Luisoni:2020efy}. Nevertheless, as we will see it is still possible to factorize the matrix elements and phase space near $\rho = \frac{1}{3}$ to isolate and extract the large logarithms, at least at the next-to-leading logarithmic level.

One may ask whether Sudakov shoulder resummation is important. For observables with only a right shoulder, such as thrust, one might argue that it is not so important, since there is not much data for $\tau>\frac{1}{3}$. However, for heavy jet mass one should generically expect that logs of the form $\alpha_s \ln^2 (\frac{1}{3} - \rho)$ are as important away from shoulder region as logs $\alpha_s \ln^2 \rho$ are away from the threshold $\rho=0$. This leaves a rather narrow range of intermediate values of $\rho$ where fixed-order perturbation theory might be trusted. Moreover, looking at Fig.~\ref{fig:thrustandhjm} it seems that the Sudakov shoulder effects on the left shoulder of heavy jet mass curve tend to pull it down (and away from thrust), so that resumming the left Sudakov shoulder might bring the curves closer together.
This difference of the left shoulder in thrust and heavy jet mass could help explain long standing discrepancies between fits for $\alpha_s$ using the two event shapes~\cite{Salam:2001bd,Chien:2010kc}.

In order to resum the Sudakov logs we first explore the regions of phase space that can contribute logarithms near the shoulder. We do this for the left shoulder of heavy jet mass in Section~\ref{sec:FOleft}. We find that the phase space near $\rho \lesssim \frac{1}{3}$ splits up into regions some of which generate large logarithms of $\frac{1}{3} - \rho$ and some of which do not. We find that all the logarithms come from regions with narrow jets in the light and heavy hemispheres. This is in contrast to threshold region, for which every allowed point of phase space near $\rho \sim 0$ can contribute logarithms of $\rho$. It is also in contrast to non-global logarithms, such as for the light jet mass. There, logarithms of the light jet mass come from regions where the heavy jet side does not have to contain only narrow jets. 

In Section~\ref{sec:SCET} we discuss the factorization of $\rho$ and $\tau$ near $\frac{1}{3}$. We find that near the shoulder region, the phase space and matrix elements both neatly factorize. This allows us to define a soft function, which along with the inclusive jet function, can be used to reproduce all the logarithms at NLO, and more generally the next-to-leading logarithmic series. In Section~\ref{sec:analysis} we analyze the resummed expression. We show that there are no non-global logarithms for the Sudakov shoulder; only regions related to the trijet configuration by soft or collinear radiation can generate the shoulder logs. We also find an unusual pole in the the resummed distribution, qualitatively similar to the Landau pole in the running coupling. Unlike the QCD Landau pole however, the singularity in the resummed heavy jet mass shoulder distribution is determined by the cusp anomalous dimension. Thus it is a kind of Sudakov Landau pole. Similar poles can be found in other observables, such as the Drell-Yan spectrum at small $p_T$~\cite{Frixione:1998dw,Becher:2010tm,Monni:2016ktx}.
We show that for the Sudakov shoulder case, the large Sudakov anomalous dimension contributing to this pole also enhances subleading power effects, making them comparable to the leading power result allowing the pole to be cancelled in the full distribution. We conclude in Section~\ref{sec:conc}.

\section{Next-to-leading order analysis \label{sec:FOleft}} 
As a first step towards understanding Sudakov shoulder logarithms, we analyze the matrix elements and phase space near the shoulder region in full QCD. We concentrate here on the  heavy jet mass for concreteness, but the same analysis works for thrust. 

At next-to leading order in QCD, there is the virtual contribution with 3 partons in the final state and a real emission contribution with 4 partons. The virtual contribution is proportional to the LO cross section and serves to regularize infrared and collinear divergences. Thus we focus on the real emission contributions  to extract the logarithms.

To have $\rho \lesssim \frac{1}{3}$ we can have configurations which differ from the trijet configuration by soft and collinear emissions, or configurations which do not. For example, one could take a  non-planar 4-parton configuration with 4 well-separated partons and $\rho \sim 0.4$, then adjust their momenta to lower $\rho$. Staring from such a configuration, one would not expect anything unusual to happen as $\rho$ is lowered through $\frac{1}{3}$. Indeed, $\rho= \frac{1}{3}$ is only special because it is a kinematic limit for 3-body phase space. Thus we expect that the only 4-parton configurations which will contribute Sudakov shoulder logarithms are those close to the trijet configuration. We will find that this is in fact the case. 

\subsection{Kinematics}
Let us define the momenta of the 4 particles in the final state as $p_1^\mu ,p_2^\mu, p_3^\mu$ and $p_4^\mu$. 
After momentum conservation, on-shell conditions and a frame choice, there are 5 independent degrees of freedom of these four momenta. 
Although we will not restrict the momenta to be
soft or collinear, it is helpful to choose variables so that the soft and collinear limits are transparent. 
To impose the on-shell constraints, it is helpful to parameterize the momenta initially in lightcone coordinates:
\begin{equation}
  p_1 = z_1 n^{\mu} + \frac{p^2_{\perp}}{4 z_1} \bar{n}^{\mu} +
  p^{\mu}_{\perp}, \quad p_2 = z_2 n^{\mu} + \frac{p^2_{\perp}}{4 z_2}
  \bar{n}^{\mu} - p^{\mu}_{\perp}
\end{equation}
\begin{equation}
  p_3 = z \omega n^{\mu} + \frac{q^2_{\perp}}{4 z \omega} \bar{n}^{\mu} +
  q^{\mu}_{\perp}, \quad p_4 = (1 - z) \omega n^{\mu} + \frac{q^2_{\perp}}{4
  (1 - z) \omega} \bar{n}^{\mu} - q^{\mu}_{\perp}
\end{equation}
where $n^\mu=(1,0,0,1)$ and $\bar{n}^\mu=(1,0,0,-1)$ are back-to-back lightlike directions. Imposing momentum conservation and defining $\phi$ as the azimuthal angle between the 1-2 and 3-4 planes we can then express all the momenta in terms of
\begin{equation}
    s_{234}=(p_2+p_3+p_4)^2,\quad s_{34}=(p_3+p_4)^2,\quad z, \quad\omega \quad \text{and} \quad \phi \label{variables}
\end{equation}
We conventionally define $\phi$ by $p_\perp^\mu = (0,p_T \sin\phi, p_T \cos\phi,0)$. The variables
$\omega = \frac{1}{2} \bar{n} \cdot(p_3+p_4)$  and $s_{234}$ are hard variables, approaching $\frac{1}{3}$ at the trijet configuration. $s_{34}$ is the invariant mass of one of the jets in the collinear limit which approaches zero in the trijet limit. The collinear momentum fraction $z$ and the azimuthal angle are order $1$ in the collinear limit, but $z\to0$ in the limit that $p_4$ is soft.
We also find it sometimes convenient to trade $\cos \phi$ for $s_{23}$ using
\begin{multline}
  s_{23} = \frac{s_{34}^2 z + 2 s_{34} (1 + 2 z \omega - 2 z - 2 \omega)
  \omega + 4 z s_{234} \omega^2 + s_{34} s_{234} (z - 1 + 2 \omega - 4 z \omega)}{4
  \omega^2 - s_{34}}\\
  + \frac{2 \sqrt{s_{34} (1 - z) z (2 \omega - s_{34}) (1 - 2 \omega) (2
  \omega - s_{234}) (2 \omega s_{234} - s_{34})}}{4 \omega^2 - s_{34} } \cos \phi
\end{multline}
When using $s_{23}$ the physical constraint $-1 \le \cos \phi \le 1$ must be imposed on the region of integration. Another useful exact relation is
\begin{equation}
    s_{12} = 1 - 2 \omega + s_{34} \left( 1 - \frac{1}{2 \omega} \right)
\end{equation}
We can use this relation to trade $\omega$ for $\rho$ when $\rho=s_{12}$.

To compute thrust or heavy jet mass, we need to determine the thrust axis from the formula in Eq.~\eqref{tdef}. With 4 partons, the two possibilities are that 3 partons are in one hemisphere and 1 parton in the other, or 2 partons can be in each hemisphere.
If we know that partons $p_1 \ldots p_m$ are to be
clustered in the same hemisphere then 
\begin{equation}
  \max_{\vec{n}} \sum_{j=1}^m | \vec{p}_j \cdot \vec{n} |
  = 2  \max_{\vec{n}} ( \sum_{j=1}^m \vec{p}_j) \cdot \vec{n}
  \label{thrustform}
\end{equation}
This dot product will be maximized if $\vec{n} =| \sum
\vec{p}_j|^{-1} \sum \vec{p}_j$ so that the thrust axis will always align with the sum of momenta in each hemisphere. So there are 7 possibilities for the thrust axis. For each axis choice
\begin{equation}
 \sum_{j=1}^m  | \vec{p}_j \cdot \vec{n} | = 2 \frac{1}{| \sum
\vec{p}_j|}  ( \sum \vec{p}_j)\cdot  ( \sum \vec{p}_j)= 2 \Big|   \sum_{j=1}^m \vec{p}_j \Big|
\end{equation}
Thus to determine the thrust axis, we need to find which set of partons has the largest value of $2 |\sum \vec{p}_j|$ or equivalently 
\begin{equation}
    T_j^2  \equiv 4|\sum \vec{p}_j|^2
\end{equation}
In terms of our variables in Eq.~\eqref{variables}, the $T_j$ with one parton in one hemisphere are relatively simple
\begin{align}
  T_1^2 &= T_{234}^2 =(1 - s_{234})^2,\quad 
  T_2^2 = T_{134}^2 =\left[\frac{2 (1 + s_{234} - 2 \omega) \omega - s_{34}}{2\omega}  \right]^2\\
  T_3^2 &= T_{124}^2= \left[\frac{4 z \omega^2 + s_{34} (1 - z)}{2\omega}
  \right]^2,\quad 
  T_4^2 = T_{123}^2 =\left[\frac{4 (1 - z) \omega^2 + s_{34} z}{2\omega}
  \right]^2
  \end{align}
We also have
\begin{equation}
  T_{12}^2 = \left( \frac{4 \omega^2 - s_{34}}{2\omega}\right)^2
\end{equation}
and
\begin{align}
    T_{13}^2&=\frac{1}{4\omega^2}\Big[s_{34}^2(1-z)^2+4\omega^2\left(4s_{23}+s_{234}^2+(1-2\omega z)^2-2s_{234}(1+2\omega z)\right)\notag\\
    &\hspace{20mm}
    -4s_{34}\omega\left(1+s_{234}-z-s_{234}z+2\omega\left(z^2-z-2\right)\right)\Big]\\
    T_{14}^2&=\Big[\frac{s_{34}z-2\omega \left(1+s_{234}-2\omega(1-z)\right)}{2\omega}\Big]^2-4s_{23}
\end{align}
All of these $T_j$ values are exact.

\begin{figure}[t]
    \centering
    \includegraphics[scale=1]{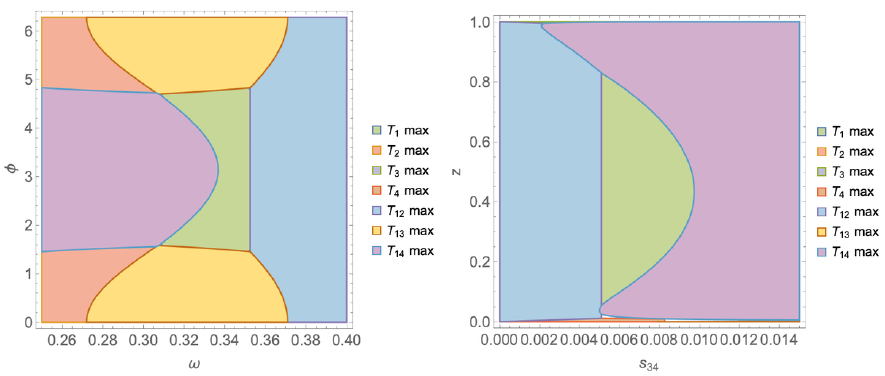}
    \caption{Example slices of 5D phase space for four massless partons. The colors indicate which collection of momenta determines the thrust axis. The left plot has $z=0.06, s_{34}=0.02$ and $s_{234} = \frac{1}{3} -0.01$. The green region in this plot would contribute to the heavy jet mass distribution at $r=\frac{1}{3}-\rho=0.01$. The right plot has $\phi=\pi$, $s_{234} = \frac{1}{3}$ and $s_{12} = \frac{1}{3} - 0.01$. The blue region in this plot contributes also at $r=0.01$.
    \label{fig:rhoregions}}
\end{figure}

Now we would like to consider the region $\rho < \frac{1}{3}$. The heavy hemisphere can have either 2 partons or 3 partons. We can therefore choose it to be $\rho = s_{234}$ with $T_1$ is maximal, or $s_{12}$ with $T_{12}$ is maximal. The other cases are given by permutation of the indices. 
Figure~\ref{fig:rhoregions} shows examples of the phase space regions labeled by which $T_j$ is greatest. All regions in these plots contribute to some value of $\rho$. However, to avoid overcounting we only need to consider the green region on the left plot and the blue region in the right plot. 

\subsection{Matrix Elements}
Let us define
\begin{equation}
    r\equiv \frac{1}{3}-\rho
\end{equation}
As we have discussed, we expect contributions to the NLO heavy jet mass cross section with factors of $\ln r$ or $\ln^2r$ to come from soft or collinear regions of phase space close to the trijet configuration. 
We can therefore power-expand the matrix elements and phase space constraints in soft and collinear limits. This dramatically simplifies the calculation. There are two ways to confirm that only soft and collinear limits are relevant. First, we can extend the integration limits to the full phase space and verify that no additional logarithms can be generated. Second we can compare the logarithms we extract with a numerical computation of the heavy jet mass distribution at NLO.

For power counting we take $r \sim \lambda \ll 1$. In the collinear limit where $p_4\, ||\, p_3$, the  phase space variables scale as
\begin{equation}
s_{34} \sim \lambda, \quad 
x \equiv \omega - \frac{1}{3} \sim \lambda, \quad
y \equiv s_{234} - \frac{1}{3} \sim \lambda, \quad
z \sim \lambda^0, \quad
\phi \sim s_{23} \sim \lambda^0
\end{equation}
In the soft limit,  where $p_3$ is soft, the scaling is the same except that $z\sim \lambda$ instead of $z \sim \lambda^0$. 

First we compute the matrix elements-squared at leading power. We do this by summing all the relevant Feynman diagrams, squaring the amplitudes and summing over spins, after which we take the leading power expansion. We cross check the results against the expectation for soft and collinear limits from factorization.

The $\gamma^\star \to q\bar{q} g$ matrix element depends on whether the gluon is polarized in the plane of scattering or out of the plane. We find
\begin{equation}
   \sum_{\text{spins}}\left|\mathcal{M}^{\text{in}}_{\gamma^{*} \rightarrow q\bar{q} g}\right|^{2}      =
   \begin{gathered}
   \resizebox{2cm}{!}{
\begin{tikzpicture}
\begin{feynman}[large]
\vertex (a); 
\vertex [position=0:1.5 from a] (b);
\vertex [position=120:1.5 from a] (c);
\vertex [position=240:1.5 from a] (d);
\diagram* {
(a) -- [gluon, edge label=\(\epsilon_\text{in}~\uparrow\),inner sep=10pt] (b),
(c) -- [fermion] (a),
(a) -- [fermion] (d),
}; 
\draw[fill=lightgray] (0,0) circle[radius=0.2];
\end{feynman}
\end{tikzpicture}
}
\end{gathered}
   =   |\cM_0|^2 2 g_s^2 C_F \label{M3pol1}
\end{equation}
when the gluon polarization $\epsilon_\text{in} = (0,0,1,0)$ in the conventions of Fig.~\ref{fig:trijet}, where $p_g = \frac{Q}{3} (1,0,0,1)$. And
\begin{equation}
   \sum_{\text{spins}}\left|\mathcal{M}^{\text{out}}_{\gamma^{*} \rightarrow q\bar{q} g}\right|^{2}      =
   \begin{gathered}
   \resizebox{2cm}{!}{
\begin{tikzpicture}
\begin{feynman}[large]
\vertex (a); 
\vertex [position=0:1.5 from a] (b);
\vertex [position=120:1.5 from a] (c);
\vertex [position=240:1.5 from a] (d);
\diagram* {
(a) -- [gluon, edge label=\(\epsilon_\text{out}~\otimes\),inner sep=10pt] (b),
(c) -- [fermion] (a),
(a) -- [fermion] (d),
}; 
\draw[fill=lightgray] (0,0) circle[radius=0.2];
\end{feynman}
\end{tikzpicture}
}
\end{gathered}
   =   |\cM_0|^2 14 g_s^2 C_F \label{M3pol2}
\end{equation}
when the gluon polarization is $\epsilon_\text{out} = (0,1,0,0)$. The sum of these agrees with Eq.~\eqref{M1form}.


The matrix elements depend on which partons are gluons and which are quarks. If $p_2$ is a quark and $p_4$ is a gluon, then to leading power in collinear scaling
\begin{equation}
   \left|\mathcal{M}^\text{collinear}_{\gamma^{*} \rightarrow q g \bar{q} g}\right|^{2} 
   =
   \begin{gathered}
   \resizebox{3cm}{!}{
\begin{tikzpicture}
\begin{feynman}[large]
\vertex (a); 
\vertex [position=-5:1.5 from a] (b1);
\vertex [position=5:1.5 from a] (b2);
\vertex [position=120:1.5 from a] (c);
\vertex [position=240:1.5 from a] (d);
\diagram* {
(a) -- [fermion, edge label=\(p_3\),inner sep=10pt] (b1),
(b2) -- [gluon, edge label=\(p_4\),inner sep=10pt] (a),
(c) -- [gluon, edge label=\(p_1\)] (a),
(a) -- [fermion, edge label=\(p_2\)] (d),
}; 
\draw[fill=lightgray] (0,0) circle[radius=0.2];
\end{feynman}
\end{tikzpicture}
}
\end{gathered}
   \cong
   \left|\mathcal{M}_{0}\right|^{2} 32 g_{s}^{4} C_{F}^{2} \frac{1}{s_{34}} \frac{1+z^2}{1-z} \label{CF2coll}
\end{equation}
Here the blob represents all the diagrams that can contribute.
We derive this by squaring the full matrix element for $\gamma^\star \to q g \bar{q} g$ using \textsc{Qgraf}\cite{NOGUEIRA1993279} and \textsc{Form}\cite{Vermaseren:2000nd} or \textsc{FeynCalc}\cite{HAHN2001418,MERTIG1991345,SHTABOVENKO2016432}, summing over spins, and then and power expanding in the small $s_{34}$ limit. The splitting function naturally appears. 

When $p_3$ and $p_4$ are gluons, then we find
\begin{align}
\left|\mathcal{M}^\text{collinear}_{\gamma^{*} \rightarrow q \bar{q} g g}\right|^{2} &=
   \begin{gathered}
   \resizebox{3cm}{!}{
\begin{tikzpicture}
\begin{feynman}[large]
\vertex (a); 
\vertex [position=-10:1.5 from a] (b1);
\vertex [position=10:1.5 from a] (b2);
\vertex [position=120:1.5 from a] (c);
\vertex [position=240:1.5 from a] (d);
\diagram* {
(a) -- [gluon, edge label=\(p_3\),inner sep=15pt] (b1),
(b2) -- [gluon, edge label=\(p_4\),inner sep=15pt] (a),
(c) -- [fermion, edge label=\(p_1\)] (a),
(a) -- [fermion, edge label=\(p_2\)] (d),
}; 
\draw[fill=lightgray] (0,0) circle[radius=0.2];
\end{feynman}
\end{tikzpicture}
}
\end{gathered}
\cong
\left|\mathcal{M}_{0}\right|^{2} C_{F} C_{A} g_{s}^{4} \frac{64}{s_{34}}\left[\frac{\left(1-z+z^{2}\right)^{2}}{z(1-z)}+\frac{3}{4} z(1-z) \cos 2 \phi\right] \label{M4qggq}
\end{align}
Note the azimuthal angle dependence is due to the polarization of the gluons. Indeed, the leading order $\gamma^\star \to q \bar{q} g$ matrix element is polarized, and we must therefore use polarized splitting functions (see~\cite{Ellis:1996mzs} for example). We have checked that summing the polarized leading order matrix elements in Eqs.~\eqref{M3pol1} and~\eqref{M3pol2} with the polarized splitting functions (see~\cite{Ellis:1996mzs} for example) reproduces Eq.~\eqref{M4qggq}.

And finally when $p_3$ and $p_4$ are quarks (or antiquarks), the leading power result is the same whether they are identical or not. 
\begin{align}
\left|\mathcal{M}^\text{collinear}_{\gamma^{*} \rightarrow q \bar{q} q^{\prime} \bar{q}^{\prime}}\right|^{2} &\cong\left|\mathcal{M}^\text{collinear}_{\gamma^{*} \rightarrow q \bar{q} q \bar{q}}\right|^{2}  
=
   \begin{gathered}
   \resizebox{3.5cm}{!}{
\begin{tikzpicture}
\begin{feynman}[large]
\vertex (a); 
\vertex [position=-10:2 from a] (b1);
\vertex [position=10:2 from a] (b2);
\vertex [position=0:1 from a] (g);
\vertex [position=120:1.5 from a] (c);
\vertex [position=240:1.5 from a] (d);
\diagram* {
(a) -- [gluon] (g),
(g) -- [fermion, edge label=\(p_3\),inner sep=15pt] (b1),
(b2) -- [fermion, edge label=\(p_4\),inner sep=15pt] (g),
(c) -- [fermion, edge label=\(p_1\)] (a),
(a) -- [fermion, edge label=\(p_2\)] (d),
}; 
\draw[fill=lightgray] (0,0) circle[radius=0.2];
\end{feynman}
\end{tikzpicture}
}
\end{gathered}
 \\
 &\cong
\left|\mathcal{M}_{0}\right|^{2} C_{F} T_f n_f g_{s}^{4} \frac{16}{s_{34}}\left(2-z-z^{2}-6 z(1-z) \cos ^{2} \phi\right) 
\label{eq:nf_matrix}
\end{align}
This expression also depends on the azimuthal angle, and like the gluon case, is consistent with using the polarized 3 parton matrix elements and polarization-dependent splitting functions. 

For the soft limits, we can power expand the full matrix elements in the soft limit. 
When $z$ is soft we cannot drop $s_{34}$ with respect to $z$, or vice-versa. When $p_3$ and $p_4$ are both gluons, the result can be written as
\begin{equation}
   \left|\mathcal{M}^\text{soft}_{\gamma^{*} \rightarrow q \bar{q} g g}\right|^{2} 
   \cong |\cM_0|^2 C_F g_s^4 \frac{64}{3} 
   \left[ \left(C_F - \frac{1}{2} C_A\right)\frac{1}{s_{14} s_{24}} +  \frac{C_A}{2}\frac{1}{s_{14} s_{34}} + \frac{C_A}{2}  \frac{1}{s_{24} s_{34}}\right]
\end{equation}
where
\begin{align}
    s_{14} s_{24} &= 9 s_{34}^2 + 16 z^2 - 24 s_{34} z \cos(2\phi) \label{s14s24}\\
    s_{14} s_{34} &= 3 s_{34}^2 + 4 s_{34} z - 4 s_{34} \sqrt{3 s_{34} z} \cos \phi\\
    s_{24} s_{34} &= 3 s_{34}^2 + 4 s_{34} z + 4 s_{34} \sqrt{3 s_{34} z} \cos \phi \label{s24s34}
\end{align}
This is consistent with the Eikonal approximation. 

To avoid double-counting we also need the soft collinear matrix elements which come from taking the soft limit (small $z$) of the collinear matrix elements or equivalently the collinear limit ($s_{34}\ll z$) of the soft matrix elements. These are therefore the same as the soft matrix elements but keeping only the final term in Eqs.~\eqref{s14s24}-\eqref{s24s34}. 

\subsection{Phase space}
For the phase space limits, we will first examine the soft-collinear limit where
$z\sim\lambda$. To leading power in the soft-collinear limit
\begin{align}
    T_1 &\cong \frac{1}{9}-\frac{y}{3}, \quad
    T_2 \cong \frac{1}{9}-\frac{s_{34}}{2}-\frac{2 x}{3}+\frac{y}{3},\quad
    T_3 \cong 0,\quad 
    T_4 \cong \frac{1}{9} + \frac{2 x}{3}-\frac{2 z}{9}\\
    T_{12} &\cong\frac{1}{9} -\frac{s_{34}}{2}+\frac{2 x}{3},\quad
    T_{13} \cong \frac{1}{9} + s_{23}-\frac{y}{3}-\frac{4 z}{9},\quad
    T_{23} \cong \frac{1}{9} - s_{23} -\frac{2 x}{3}+\frac{y}{3}+\frac{2 z}{9}
\end{align}

For the case where $T_1$ is maximal, $\rho = s_{234}$ and $y=r$. We can then impose the constraints $T_1>T_2$, $T_1 > T_3$, and so on. Since we are using the variable $s_{23}$ instead of $\cos \phi$ we also have to impose $-1\le \cos \phi \le 1$. Reducing these constraints leads to five integration regions
\begin{equation}
\begin{gathered}
\int d\Pi_1=2 \int_{0}^{\frac{1}{2}s_{34}^{+}} d s_{34}\int_{z^{+}}^{1-z^{+}} dz \int_{x_A}^{x_B} d x \int_{s_{23}^{\phi-}}^{s_{23}^{\phi+}} d s_{23}\ J+2 \int_{0}^{s_{34}^{+}} d s_{34}\int_{\frac{9}{4} s_{34}}^{z^{+}} dz \int_{x_C}^{x_B} d x \int_{s_{23}^{\phi-}}^{\frac{4}{9} z} d s_{23}\ J \\
+2 \int_{0}^{s_{34}^{+}} d s_{34}\int_{\frac{9}{4} s_{34}}^{z^{+}} dz \int_{x_A}^{x_C} d x \int_{s_{23}^{A}}^{\frac{4}{9} z} d s_{23}\ J+2 \int_{0}^{s_{34}^{+}} d s_{34} \int_{z^{-}}^{\frac{9}{4} s_{34}} dz\int_{x_C}^{x_D} d x \int_{s_{23}^{\phi-}}^{\frac{4}{9} z} d s_{23}\ J \\
+\int_{0} d s_{34}\int_{z^{-}}^{\frac{9}{4} s_{34}} dz \int_{x_E}^{x_C} d x \int_{s_{23}^A}^{\frac{4}{9} z} d s_{23}\ J
\end{gathered}
\end{equation}
where 
\begin{align}
s_{34}^{+} &=\frac{4}{9}(7-4 \sqrt{3}),\quad
z^{\pm}  =\frac{9}{4}(7 \pm 4 \sqrt{3}) s_{34} \\
s_{23}^{\phi \pm}&=\left(\sqrt{\frac{s_{34}}{4}} \pm \sqrt{\frac{z}{3}}\right)^{2}, \quad 
s_{23}^{A} =-\frac{2}{3} x+\frac{2}{3} y+\frac{2}{9} z \\
x_A & =-\frac{3}{4} s_{34}+y, \quad 
x_B =\frac{3}{4} s_{34}-\frac{y}{2}, \quad x_C=-\frac{3}{8} s_{34}+y+\frac{\sqrt{3 s_{34} z}}{2}-\frac{z}{6} \\
x_D &=-\frac{y}{2}+\frac{z}{3}, \quad x_E=y-\frac{z}{3}
\end{align}
The Jacobian
\begin{equation}
    J
    =\frac{1}{\sqrt{(s_{23}^{\phi+}-s_{23})(s_{23}-s_{23}^{\phi-})}}
\end{equation}
scales like $J\sim \lambda^0$.
For the leading double log we need to compute 
\begin{equation}
    \int d\Pi_1 |\cM|^2 \sim \int d\Pi_1 \frac{1}{s_{34} z}
\end{equation}
Analyzing the integrals we find that none of them generate $\ln r$ terms; the limit $r\to 0$ in each of the integrals is smooth. Thus the region with $T_1$ max does not contribute to the Sudakov shoulder at NLO. The logs must therefore come from regions with two partons in each hemisphere. 

Next, we consider configurations where $T_{12}$ is maximal. As before, we expand first assuming collinear scaling. In this case, we no longer have $r=\frac{1}{3}-\rho = y$ but instead 
\begin{equation}
    \rho = s_{12} = 1 + s_{34} - \frac{s_{34}}{2 \omega} - 2 \omega \approx \frac{1}{3} - \frac{1}{2} s_{34} - 2 x
\end{equation}
So that $r=\frac{1}{2} s_{34} + 2 x$. To hold $r$ fixed we then can use $r, s_{34}, z, y, s_{23}$ as independent variables (instead of $r,s_{34},z,x, s_{23}$ in the $T_1$ max case). 
Now we find 40 relevant integration regions. In most of these $r$ can be set to zero without consequence. Only four can possibly generate logs of $r$:
\begin{multline}
\label{P12int}
    \int d \Pi_{12} =
    \int_0^r d s_{34} \int_{z^+}^{1-z^+} d z \int_{y_A}^{y_B} dy
    \int_0^{2\pi} d\phi 
    + \int_0^{\frac{r}{3}} d s_{34} 
    \int_{\frac{9 s_{34}}{4}}^{z^+} d z \int_{y_D}^{y_C} dy
    \int_0^{2\pi} d\phi 
    \\
+ \int_0^{\frac{r}{3}} d s_{34} 
    \int_{\frac{9 s_{34}}{4}}^{z^+} d z \int_{y_C}^{y_B} dy \int_{s_{23}^B}^{s_\phi^+} d s_{23}\ J
     + \int_0^{\frac{r}{3}} d s_{34} 
    \int_{\frac{9 s_{34}}{4}}^{z^+} d z \int_{y_A}^{y_D} dy \int_{s_{23}^{\phi-}}^{s_{23}^C} d s_{23}\ J 
\end{multline}
where
\begin{align}
 y_A &= - r + 2 s_{34}, \quad y_B = 2 r - s_{34},\notag\\
  y_C &=
   2 r - \frac{7 s_{34}}{4} - \sqrt{3 s_{34} z} + \frac{z}{3}, \quad y_D = - r
   + \frac{11 s_{34}}{4} + \sqrt{3 s_{34} z} - \frac{z}{3} \notag \\
 s_{23}^B &= -
  \frac{2 r}{3} + \frac{5 s_{34}}{6} + \frac{y}{3} + \frac{2 z}{9}, \quad
  s_{23}^C = \frac{r}{3} - \frac{2 s_{34}}{3} + \frac{y}{3} + \frac{4 z}{9}
\end{align}

For the $C_F^2$ color structure, using the power-expanded matrix elements in the collinear limit, Eq.~\eqref{CF2coll}, only the first two integrals in Eq.~\eqref{P12int} contribute. We find 
\begin{equation}
     \mathcal{S}_{\text{c}}^{(C_F)}=4 \int d\Pi_{12} |\mathcal{M}_{\gamma^*\to q g g \bar q}^{\text{collinear}}|^2 =\frac{6\alpha_s^2}{\pi^2}C_F^2 r\left[-2\ln^2 r+\left(1-8\ln\frac{3}{2}\right)\ln r+\cdots\right]
\end{equation}
Similarly, integrating against the soft matrix element and the soft-collinear overlap region, we find
\begin{align}
    \mathcal{S}_{\text{s}}^{(C_F)}&=\int d\Pi_{12} \left(4|\mathcal{M}_{\gamma^*\to q g g \bar q}^{\text{soft}}|^2+2 |\mathcal{M}_{\gamma^*\to q \bar q g g}^{\text{soft}}|^2\right)=\frac{12\alpha_s^2}{\pi^2}C_F^2 r\left[-\ln^2 r+2\left(1+\ln\frac{4}{3}\right)\ln r+\cdots\right]\\
    \mathcal{S}_{\text{sc}}^{(C_F)}&=4 \int d\Pi_{12} |\mathcal{M}_{\gamma^*\to q g g \bar q}^{\text{soft-coll}}|^2 =\frac{12\alpha_s^2}{\pi^2}C_F^2 r\left[-\ln^2 r+2\left(1-2\ln\frac{3}{2}\right)\ln r+\cdots\right]
\end{align}
The constants in the integrals come from the permutations of final state particles and we have accounted the symmetry factor for identical gluons. The total is
\begin{align}
    \frac{1}{\sigma_0}\frac{d\sigma^{(C_F)}}{dr}&=\mathcal{S}_{\text{c}}^{(C_F)}+\mathcal{S}_{\text{s}}^{(C_F)}-\mathcal{S}_{\text{sc}}^{(C_F)}\notag\\*
    &=\left(\frac{\alpha_s}{4\pi}\right)^2C_F^2 r\left[-192\ln^2 r +\left(96+768\ln 2-384 \ln 3\right)\ln r+\cdots\right]
\end{align}
This is compared to the exact (numerical) NLO calculation in the shoulder region in Fig.~\ref{fig:focompare}. 

For the $C_F C_A$ color structure, there can be single logarithms coming from both the the $z\sim 0$ and $z\sim 1$ regions. Moreover, the splitting functions in this case depend on the polarization of the gluon that splits. However, because the only integration regions that contribute logarithms are uniform in $\phi$ (the first two in Eq.~\eqref{P12int}), one can simply azimuthally-average the splitting functions, reducing them to the unpolarized case. The final resumts we find are we find
\begin{align}
    \mathcal{S}_{c}^{(C_A)}&=\frac{\alpha_s^2}{\pi^2}C_F C_A r \left[-6 \ln^2 r+\left(1-24\ln\frac{3}{2}\right)\ln r+\cdots\right]\notag\\
    \mathcal{S}_{s}^{(C_A)}&=6\frac{\alpha_s^2}{\pi^2}C_F C_A r \left[-\ln^2 r +2\left(1+\ln \frac{4}{3}\right)\ln r+\cdots\right]\notag\\
    \mathcal{S}_{sc}^{(C_A)}&=6\frac{\alpha_s^2}{\pi^2}C_F C_A r\left[-\ln^2 r +2\left(1-2\ln \frac{3}{2}\right)\ln r+\cdots\right]
\end{align}
which gives
\begin{align}
    \frac{1}{\sigma_0}\frac{d\sigma^{(C_A)}}{dr}&=\left(\frac{\alpha_s}{4\pi}\right)^2 C_F C_A r \left[-96 \ln^2 r+\left(16+384\ln 2-192 \ln 3\right)\ln r+\cdots \right]
\end{align}
Again, this is compared to NLO in the shoulder region in Fig.~\ref{fig:focompare}. 

The $n_f T_F C_F$ color structure only contains a single logarithm since there is no soft region. Integrating the collinear matrix element Eq. (\ref{eq:nf_matrix}) over power expanded phase space gives 
\begin{align}
    \frac{1}{\sigma_0}\frac{d\sigma^{(n_f)}}{dr}&=64\left(\frac{\alpha_s}{4\pi}\right)^2C_F T_f n_f r \ln r+\cdots
\end{align}
No overlap subtraction is needed. This is also shown in Fig.~\ref{fig:focompare}. 

One can perform a similar leading-power computation for the right shoulder for thrust and heavy jet mass. For these cases, we find it is only the phase space regions with 1 parton in one hemisphere and 3 partons in the other hemisphere that contribute. Since the equivalent calculation is significantly easier using Soft-Collinear Effective Theory, we skip the details of the right-shoulder cases using the full theory and turn instead to the effective theory approach.


\section{Factorization and Resummation \label{sec:SCET}}

In Section~\ref{sec:FOleft}, we computed the Sudakov shoulder logs for heavy jet mass and thrust at NLO using full QCD expanded to leading power. We now want to generalize the analysis to all orders leading to a factorization formula. To do so, we first review the approach of~\cite{Catani:1997xc} and  and discuss recoil sensitivity. We then demonstrate a different approach inspired by the NLO calculation that leads to a systematically improvable factorization formula. 

\subsection{Recoil sensitivity}
One approach to resummation of Sudakov shoulders~\cite{Catani:1997xc} is that emissions from one of the hard partons will cause an additive shift in heavy jet mass (or thrust) from $\rho \to \rho + m^2$. Then one could write  the resummed distribution as a convolution. Heuristically,
\begin{equation}
    \sigma_{\text{resummed}}(\rho) \sim \int d m^2 \sigma_{\text{LO}}(\rho -m^2) J(m^2)
    \label{convoguess}
\end{equation}
With $J(m^2)$ representing some sort of jet function and $\sigma_{\text{LO}}(\rho)$ the leading order cross section.

Unfortunately, when one tries to make this formula more precise it produces ambiguities beyond the leading logarithmic order. To see this, consider how $\rho$ changes due to emissions in the light hemisphere making the light hemisphere have a mass $m^2$. 
With 3 massless partons taking $p_1$ and $p_2$ in the heavy hemisphere and $p_3$ in the light hemisphere for concreteness, the heavy jet mass is
\begin{equation}
    \rho = (p^{\mu}_1 + p^{\mu}_2)^2 = (p^{\mu}_{\text{tot}} - p^{\mu}_3)^2 = 1 -2 E_3
    \label{rhoofE3}
\end{equation}
with $E_3$ the energy of the light-hemisphere parton. Now say the $p_3$ parton becomes massive (i.e. turns into a jet) with $p_3^2 = m^2$. Then we have the exact relation
\begin{equation}
    \rho = (p^{\mu}_1 + p^{\mu}_2)^2 = (p^{\mu}_{\text{tot}} - p^{\mu}_3)^2 =  1 + m^2 -2 E_3
\end{equation}
So it seems $\rho \to \rho + m^2$, as in Eq.~\eqref{convoguess}. However, this was a little too quick. For suppose instead of expressing $\rho$ in terms of $E_3$ we expressed it in terms of $|\vec{p}_3|$. Then when $p_3$ is massless,
\begin{equation}
    \rho =  1 -2 |\vec{p}_3|
\end{equation}
However after the emissions,
\begin{equation}
    \rho = 1+m^2 -2 E_3 =1+ m^2 -2\sqrt{\vec{p}_3^{\,2}+m^2} \cong 1 + m^2 -2 |\vec{p}_3| - \frac{m^2}{|\vec{p}_3|} + \cdots 
\end{equation}
Now, near threshold $|\vec{p}_3| \sim \frac{1}{3}$, so $\frac{m^2}{|\vec{p}_3|} \cong 3 m^2$ and we find
 $\rho \to \rho -2m^2$ instead of $\rho \to \rho + m^2$. Thus the way $\rho$ shifts depends on whether we hold the energy or the momentum of the jet fixed after the emission. This recoil-sensitivity seems to violate factorization. 
 Moreover, if $\rho\rightarrow \rho-2m^2$ one cannot write down a convolution for the distribution as in \eqref{convoguess}, since the shift implies that emissions only decrease the value of the heavy jet mass. Thus it becomes clear that while one might use the emission picture for the double-logarithmic analysis of~\cite{Catani:1997xc}, it is inadequate for NLL resummation.
 
\subsection{Factorization \label{sec:factorization}}
To proceed, recall from Section~\ref{sec:FOleft} which configurations contributed to the NLO logs. With 4 partons, we can have either 2 in each hemisphere or 1 in one light hemisphere and 3 in the heavy hemisphere. For the left shoulder of heavy jet mass at NLO we found that only the case with 2 partons in each hemisphere contributed. Moreover, the two partons in the heavy hemisphere were hard, with invariant mass $\rho \sim \frac{1}{3}$, while the two partons in the light hemisphere formed a jet of small invariant mass, $s_{34} \sim \frac{1}{3} - \rho \ll 1$. In contrast, for the right shoulder of heavy jet mass or thrust, only the region with 1 parton the light hemisphere contributed. Moreover, the configuration in the heavy hemipshere had two hard partons and one parton which was soft or collinear to one of the hard partons.

In the $r=\frac{1}{3}-\rho \ll 1$ region, we found integrals like
\begin{align}
I &\sim  |\cM_0|^2\frac{\alpha_s}{4\pi}C_F^2  \int_0^r \frac{d s_{34}}{s_{34}}
    \int_{\frac{9}{4} s_{34}}^1 \frac{dz}{z} \int_{\frac{1}{3}-r}^{\frac{1}{3} + 2r} d s_{234}\\
    &\cong |\cM_0|^2\frac{\alpha_s}{4\pi}C_F^2 r \ln^2 r
\end{align}
The integrals over $s_{34}$ (the invariant mass of the $34$ jet) and $z$ (the collinear splitting fraction in the $34$ jet) are similar to what we would have in an inclusive jet function. The $s_{234}$ variable is a hard phase space variable, equal to $s_{23}$ at leading power. The last integral gives the factor of $r$ which is the same factor in the leading order cross section, as in Eq.~\eqref{lothrust}. Thus at higher orders it is natural to expect the generalization of this integral to one with a single integral over hard kinematic phase space and an integral over the kinematics of the light jet. Thus, instead of convolution of the hard cross section with the emission cross section, as in ~\eqref{convoguess}, we should expect the phase space to factorize into a part which depends on the hard kinematics and a part which depends on the emissions.

The first observation allowing us to factorize the cross section in the region $r=\frac{1}{3}-\rho \ll 1$ is that only configurations which differ from the trijet configuration by soft or collinear emissions can generate logarithms of $r$. The reason for this is that $r=0$ is only special from the point of view of 3-body massless kinematics. One can have 4-parton configurations with $\rho$ close to $\frac{1}{3}$ that are not close to the trijet configuration. However, such configurations contribute to the cross section both for $r<0$ and $r>0$ and will be smooth across $r=0$. Hence they cannot produce large logarithms (in Section~\ref{sec:ngls} we use this same argument to show there are no non-global logs in the Sudakov shoulders). 

So let use consider a generic configuration with 3 jets pointing in the $n_1$, $n_2$ and $n_3$ directions. Such a configuration can have particles collinear to the 3 directions as well as soft partons scattered throughout phase space. At leading power, we can treat the collinear radiation as generating masses $m_1$, $m_2$ and $m_3$ for the three jets. Thus we can approximate the state as having three hard, massive particles with momenta $p_1,p_2$ and $p_3$ and soft radiation.

To compute heavy jet mass and thrust, we need to know which direction the thrust axis points for a given amount of collinear and soft radiation. To determine this, we first observe that as in Eq.~\eqref{thrustform}, the thrust axis is determined by the set of momenta in a given hemisphere that maximize
\begin{equation}
    T_{\{p_i\}} =  2\Big|\sum_{p_i} \vec{p} \,\,\Big| \label{Tform}
\end{equation}
Then $\tau$ and $\rho$ can be computed from the set $\{p_i\}$. 

Let us begin with the case where there is only collinear momenta, so we only have the 3 massive momenta to consider. In this case, phase space is described by $s_{12}, s_{13}$ and $s_{23}$ subject to $s_{12} + s_{13} +s_{23} = 1 + m_1^2+m_2^2+m_3^2$, where $s_{ij} = (p_i+p_j)^2$. Then
\begin{equation}
    T_1^2 = 4 \,\vec{p}_1^{\,2}
    =(1-s_{23})^2-2m_1^2(1+s_{23})+m_1^4 
\end{equation}
and similarly for $T_2^2$ and $T_3^2$ by permutation. 
Let us take the case where $T_1$ sets the thrust axis, so that $r = \frac{1}{3} - s_{23}$. 
Then at leading power (assuming $m_i^2 \sim r\sim s_{12} - \frac{1}{3}$)
\begin{equation}
    T_1 \cong   \frac{2}{3} + r -2 m_1^2
    \quad T_2 \cong \frac{1}{3} -r+ s_{12} -m_1^2 - 3 m_2^2 - m_3^2
    \quad T_3 \cong 1-s_{12} -  2 m_3^2
        \label{T123}
\end{equation}
So the conditions $T_1>T_2$ and $T_1>T_3$ imply
\begin{equation}
   \frac{1}{3}- r + 2 m_1^2 -2 m_3^2  < s_{12} <\frac{1}{3} +2r - m_1^2 + 3 m_2^2 + m_3^2
\end{equation}
These limits on $s_{12}$ pinch off when $r = m_2^2+m_3^2-m_1^2$. 
At $m=0$ the linear scaling with $r$ of the $s_{12}$ integration region is what generates the linear fall off of the thrust or heavy jet mass cross section as in Eq.~\eqref{lothrust}.
For the integration region to be nonzero we therefore have
\begin{equation}
    m_1^2 < r +  m_2^2+m_3^2  \label{m1bound}
\end{equation}
In other words, at fixed $m_2, m_3$ and $r$, there is an upper limit on the light-hemisphere jet mass. The probability of finding a light jet of mass at most $m_1$ at leading power is proportional to $\ln^2 m_1$, so for $m_2=m_3=0$ the integral over $m_1$ up to $r$ will give the $\ln^2 r$ left Sudakov shoulder logarithms. Combined with the factor of $r$ from the $s_{12}$ integration gives an overall $r \ln^2 r$ behavior. 
If $m_2$ and $m_3$ are parametrically larger than $r$ then we can drop $r$ in Eq.~\eqref{m1bound}. In that case, no logs are generated.  Thus the shoulder logs are determined by the region of small $m_1$, $m_2$ and $m_3$, consistent with a global observable. 

 The right shoulder for heavy jet mass is constrained by Eq.~\eqref{m1bound}, but with $r<0$. For the right shoulder we define $s=\rho-\frac{1}{3}$. Then
\begin{equation}
    m_1^2 + s < m_2^2 + m_3^2
\end{equation}
replaces Eq.~\eqref{m1bound}.

For the thrust case, we define $t=\tau-\frac{1}{3}$. When $T_1$ determines the thrust axis, then at leading power
\begin{equation}
    t = \frac{2}{3}- T_1  \cong 2m_1^2 - r
\end{equation}
and Eq.~\eqref{m1bound} becomes
\begin{equation}
    t < m_1^2 +  m_2^2+m_3^2  \label{tbound}
\end{equation}
Thus the right shoulder for thrust is defined by integrals over any of the masses with a lower limit of $t$. Since the inclusive integral, without this constraint, has no $t$ dependence, one can equivalently get the right Sudakov shoulder logarithms  by integrating over the masses constrained by $m_1^2 +  m_2^2+m_3^2< t$. 

For the soft radiation, we first need to determine when it affects the thrust axis.
Let's start with the configuration with 3 massive partons and suppose some soft radiation $k$ enters hemisphere $1$. We want to know whether the thrust axis should shift so that hemisphere $1$ excludes $k$ or if it should stay fixed, to include $k$. 
To find out, we need to compare $T_{1k}$, the thrust value with $p_1$ and $k$ included in the hemisphere, to $T_1$ where $k$ is not the 1-hemisphere, but is still included overall.
A quick calculation shows that 
\begin{equation}
    T_{1k}^2  \cong T_1^2 + \frac{8}{3}( p_2 \cdot k +  p_3 \cdot k - 2 p_1 \cdot k)
\end{equation}
Defining $p_{\bar{j}}$ as $p_j$ with its 3-momentum reversed, so
\begin{equation}
p_{\onebar}  = \frac{Q}{3}(1,0,0,-1) = \frac{2}{3} p_2 + \frac{2}{3}p_3 -\frac{1}{3} p_1 
\end{equation}
we can write
\begin{equation}
        T_{1k}^2 \cong T_1^2 + 4( p_\onebar \cdot k - p_1\cdot k)
\end{equation}
When $k$ is in the $1$ hemisphere it must be closer to $p_1$ than $p_\onebar$. In that case $p_\onebar \cdot k > p_1\cdot k$. We conclude that thrust is maximized when all the soft radiation in the hemisphere centered on $p_1$ is included.  In other words, if radiation is slightly on the opposite side of the hemisphere boundary, the thrust axis should not shift to cluster $k$ with $p_1$. 

\begin{figure}
    \centering
    \includegraphics[scale=1]{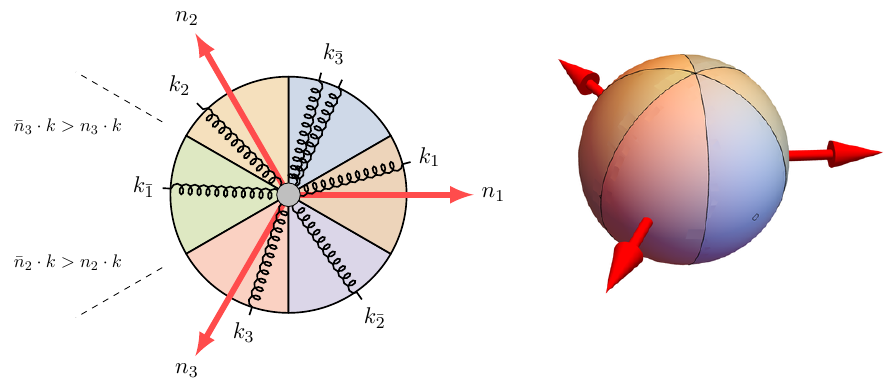}
    \caption{Soft radiation from the trijet configuration can be categorized as entering one of 6 sextant wedges shaped like carpels of an orange. The boundary of each sextant is determined by two planes orthogonal to the jet directions $n_1$, $n_2$ and $n_3$. For example, radiation in the sextant labeled $k_{\bar{1}}$ (backwards to the $1$-jet)
    is characterized by $\bar{n}_2\cdot k > n_2\cdot k$ and $\bar{n}_3\cdot k > n_3\cdot k$. 
    }
    \label{fig:oranges}
\end{figure}

Now suppose there is a lot of soft radiation with momenta with  $\{k_i^\mu\}$. Since the thrust value goes up when radiation is included in a given hemisphere, to find the thrust axis we only have to consider 3 sets of momenta: for each $j$ the set includes a hard jet's momentum $p_j$ and all the soft radiation $k_j^\hemi$ in the jet's hemisphere. That is, the maximal value of thrust for a hemisphere containing $p_j$ will be given by
\begin{equation}
    T_{j}^{\text{max}} = T_j + 
    3( p_{\bar{j}} \cdot k_j^\hemi - p_j\cdot  k_j^\hemi)
\end{equation}
Since the jet hemispheres overlap, there will be some soft radiation included in both $k_1^\hemi$ and $k_2^\hemi$, for example. To avoid overcounting, let us decompose the soft momenta into 6 regions, as shown in Fig.~\ref{fig:oranges}. So
\begin{equation}
    k_1^\hemi = k_1 + k_\twobar + k_\threebar
\end{equation}
and so on. Here $k_j$ is the soft radiation in the sextant centered on $p_j$ and $k_{\bar{j}}$ is the soft radiation in the sextant opposite to $p_j$. 

Assuming $p_1$ is the thrust axis, then heavy jet mass 
\begin{equation}
    \rho = (p_2+p_3 + k_\onebar + k_2 + k_3)^2  
\end{equation}
For $\rho<\frac{1}{3}$ we want to express constraints in terms of $r=\frac{1}{3}-\rho$. For the hard kinematic variable, we can use anything equal to  $s_{12}$ at leading power. A convenient choice is
\begin{multline}
    \xi \equiv s_{12} - \frac{1}{3} + r - 2 m_1^2 + 2 m_3^2 \\
    +2(p_2-p_1)\cdot(k_1+k_\threebar)+ 2k_2\cdot(p_1+p_2)  -4k_\twobar\cdot(p_1-p_3)+4k_3\cdot p_3 + 4 k_\onebar\cdot p_3
\end{multline}
The variable $\xi$ is defined so that $T_{1k}^{\text{max}} =T_{2k}^\text{max}$ at $\xi=0$. Then phase space where 
$T_{1k}^{\text{max}} =T_{2k}^\text{max}$ and $T_{1k}^{\text{max}} =T_{3k}^\text{max}$ is $0\le \xi \le \frac{1}{3} W$ where
\begin{equation}
    W(r,m_j,k_i) = r - m_1^2 + m_2^2  +  m_3^2 - 2 p_1 k_1 + 2 p_2 k_2 +2 p_3 k_3 
        + 2 v_\onebar k_\onebar - 2  v_\twobar k_\twobar - 2 v_\threebar k_\threebar \label{Rdef}
\end{equation}
with
\begin{align}
    v_\onebar &= -\frac{1}{3} p_1 + \frac{2}{3} p_2 + \frac{2}{3} p_3= \frac{Q}{3}(1,0,0,-1) \label{vb1}\\
    v_\twobar &= \frac{4}{3} p_1 +\frac{1}{3} p_2 - \frac{2}{3} p_3 =\frac{Q}{3} \left(1,0,\frac{\sqrt{3}}{2},\frac{3}{2}\right) \label{vb2}\\
    v_\threebar &= \frac{4}{3} p_1 - \frac{2}{3} p_2 + \frac{1}{3} p_3 =\frac{Q}{3}\left(1,0,-\frac{\sqrt{3}}{2},\frac{3}{2}\right)\label{vb3}
\end{align}
We have fixed the signs of the $v_{\bar{j}}$ so that they all have positive energy. Since $v_\onebar = p_\onebar$, and $k_\onebar$ is close to $p_\onebar$, we will have $v_\onebar \cdot k_\onebar\ge 0$ for all $k_\onebar$. For the other directions, $\vec{v}_\twobar \cdot \vec{p}_2 =0$ and $\vec{v}_\threebar \cdot \vec{p}_3 =0$, and they will also have $v_\twobar \cdot k_\twobar\ge 0$ and  $v_\threebar \cdot k_\threebar\ge 0$.  

For the integration range over $\xi$ to be nonzero we therefore need
\begin{equation}
{
 m_1^2 + 2 p_1 k_1 + 2  v_\twobar k_\twobar +2  v_\threebar k_\threebar < r  + m_2^2 + 2 p_2 k_2 + m_3^2 + 2 p_3 k_3 + 2v_\onebar k_\onebar \label{rform}
}
\end{equation}
which is the same as $W(r,m_j,k_i)>0$, with $W$ in Eq.\eqref{Rdef}.
Every term in this expression is a positive quantity. This inequality applies to both the left and right shoulder for heavy jet mass (for the right shoulder we prefer to use $s=-r=\rho-\frac{1}{3}>0$).

For thrust, defining $t=\tau - \frac{1}{3} = \frac{2}{3}- T_{1k}$ the bound is $0 \le x \le T$ where
\begin{equation}
{
    T(t,m_j,k_i) = m_1^2 +m_2^2 + m_3^2 + 2 p_1 k_1 + 2 p_2 k_2 + 2 p_3 k_3+ 2v_\onebar k_\onebar  
    + 2  v_\twobar' k_\twobar +2  v_\threebar' k_\threebar -t \label{Tdef}
}
\end{equation}
where
\begin{align}
    v_\twobar' = 2p_1-v_\twobar = \frac{Q}{3}\left(1,0,-\frac{\sqrt{3}}{2},\frac{1}{2}\right)=\frac{Q}{3}\bar{n}_2  \label{vbars1}\\
    v_\threebar' = 2p_1-v_\threebar = \frac{Q}{3}\left(1,0,\frac{\sqrt{3}}{2},\frac{1}{2}\right)=\frac{Q}{3}\bar{n}_3 \label{vbars2}
\end{align}
So that
\begin{equation}
    t< m_1^2 +m_2^2 + m_3^2 + 2 p_1 k_1 + 2 p_2 k_2 + 2 p_3 k_3+ 2v_\onebar k_\onebar  
    + 2  v_\twobar' k_\twobar +2  v_\threebar' k_\threebar 
    \label{tform}
\end{equation}
For thrust, as for heavy jet mass, every term in this inequality is positive.


As observed in Section~\ref{sec:FOleft}, we can set $m_j=k_i=0$ to zero in the hard matrix elements at leading power. Then the integral over hard phase space simply gives the maximum value of $\xi$ from Eqs.~\eqref{Rdef} or \eqref{Tdef}. That is, each channel of the LO integral in Eq.~\eqref{lothrust} gets modified as
\begin{equation}
 48 C_F\frac{\alpha_s}{4\pi} \int_0^{\frac{1}{3}-\tau} d s_{12}
    \to  
     48 C_F\frac{\alpha_s}{4\pi} \int_0^{R/3} d \xi =  48 C_F\frac{\alpha_s}{4\pi} R\, \theta(R)
\end{equation}
with $\theta(x)$ the Heaviside step function. 

The rate for producing collinear radiation is given by splitting functions, and the cross section for producing collinear radiation of mass $m$ is given by the inclusive jet function $J(m^2)$. 
The rate for soft radiation is given by a soft function, defined as an integral over emissions from Wilson lines using a measurement function (see Section~\ref{sec:soft}). The key equation, Eq.~\eqref{rform}, lets us then write the factorized expression for the heavy jet mass Sudakov shoulder as
\begin{equation}
   \frac{1}{\sigma_1} \frac{d \sigma}{d r} = H(Q) \int d^3 m^2 d^6 q J(m_1^2)J(m_2^2) J(m_3^2) S_6(q_i) W(m_j,q_i,r) \theta[\,W(m_j,q_i,r) ]
\end{equation}
where
\begin{equation}
    \sigma_1 =   48 C_F\frac{\alpha_s}{4\pi} \sigma_0
    \label{sigma1def}
\end{equation}
The arguments of the 6-parameter soft function $S_6(q_i)$ are the projections $q_i = n_i \cdot k_i$ and $q_{\bar{i}} = v_{\bar{i}} \cdot k_{\bar{i}}$. In terms of the $q_i$, Eq. \eqref{Rdef} becomes
\begin{equation}
    W(m_j,q_i,r) 
    = r - m_1^2 + m_2^2  +  m_3^2 +\frac{2Q}{3} (q_2+q_3+q_\onebar - q_1 -q_\threebar-q_\twobar) \label{Rdef2}
\end{equation} 
We can simplify the factorized expression by defining a 2-parameter trijet hemisphere soft function
\begin{equation}
    S(q_\ell, q_h) = \int d^6 q_i\ S_{6}(q_i)
    \delta(q_\ell - q_1 - q_\twobar - q_\threebar)
    \delta(q_h - q_\onebar - q_2 - q_3)
    \label{trijethemi}
\end{equation}
where $q_\ell$ and $q_h$ represent the soft radiation in the light and heavy hemispheres. This soft function contributes to the doubly-differential distribution of the hemisphere masses as 
\begin{multline}
    \frac{d^2\sigma}{d m_\ell^2 d m_h^2} = 
   H(Q,\mu) \int d m_1^2\ d m_2^2\ d m_3^2\ d q_\ell\ d q_h\  J(m_1^2,\mu)J(m_2^2,\mu)J(m_3^2,\mu) S(q_\ell,q_h,\mu)\\ 
    \times  \delta\left(m_\ell^2-m_1^2 -\frac{2}{3} q_\ell Q\right) \delta\left(m_h^2 - m_2^2 -m_3^2 -\frac{2}{3} q_h Q\right) \label{doublesig}
\end{multline}
And then
\begin{equation}
    \frac{d\sigma}{dr} = \int dm_h^2 dm_\ell^2 \frac{d^2\sigma}{d m_\ell^2 d m_h^2} (r+m_h^2-m_\ell^2) \Theta(r+m_h^2-m_\ell^2)
    \label{dsdr1}
\end{equation}
One also must sum over channels, corresponding to which jet is the quark jet, which is antiquark and which is gluon.

 The factorization formula for thrust is similar: 
 \begin{equation}
    \frac{d\sigma}{dt} = \int dm_h^2 dm_\ell^2 \frac{d^2\sigma}{d m_\ell^2 d m_h^2} (m_h^2+m_\ell^2-t) \Theta(m_h^2+m_\ell^2-t) \label{dsdr2}
\end{equation}
The soft function for thrust is the same as for heavy jet mass after changing $v\to v'$. As we will show in Appendix~\ref{appendix:soft} changing $v\to v'$ has no effect on the parts of the soft function relevant to NLL resummation, so we will treat the HJM and thrust trijet hemisphere soft functions as being the same. 

 \subsection{Soft function \label{sec:soft}}
According to the analysis in the previous section, the factorization formula requires a soft function giving the rate for producing gluons $k_i$ entering one of 6 sextants, as in Fig.~\eqref{fig:oranges}. In each sextant we need the projection $p_i\cdot k_i$ for $i=1,2,3$ (sextants containing a jet) or $v_i \cdots k_i$ for $i=\onebar,\twobar,\threebar$ (sextants between jets). For NLL resummation, we only need the anomalous dimension of the soft function at 1-loop. This can be determined by RG invariance. However, as a cross check on the factorization formula, it is important to compute the soft function explicitly. 

It is convenient to introduce the scaleless vectors for the six direction that appear in the measurment function
\begin{equation}
    p_i = \frac{Q}{3}n_i,\quad v_{\ibar} = \frac{Q}{6} N_i,\quad i=1,2,3
\end{equation}
where the $n_i$ can be read off from  Fig.~\ref{fig:trijet} and the $N_i$ from Eqs.~\eqref{vb1}-\eqref{vb3} (or see Appendix~\ref{appendix:soft}). The vectors $p_i$, $n_i$, $v_\onebar$ and $N_\onebar$ are lightlike while $v_\twobar, v_\threebar, N_\twobar$ and $N_\threebar$ are spacelike. 
For heavy jet mass, the measurement function $M(k,q_i)$ is
\begin{align}
    M(k,q_i) &= 
    \theta(n_\twobar \cdot k - n_2\cdot k) \theta(n_\threebar \cdot k - n_3\cdot k)\delta\left(q_1-\frac{2}{3} n_1\cdot k\right)\\*
    &+
     \theta(n_\threebar \cdot k - n_3\cdot k) \theta(n_\onebar \cdot k - n_1\cdot k)\delta\left(q_2- \frac{2}{3} n_2\cdot k\right)\\*
         &+
     \theta(n_\onebar \cdot k - n_1\cdot k) \theta(n_\twobar \cdot k - n_2\cdot k)\delta\left(q_3-\frac{2}{3} n_3\cdot k\right)\\*
         &+
     \theta(n_2 \cdot k - n_\twobar\cdot k) \theta(n_3 \cdot k - n_\threebar\cdot k)\delta\left(q_\onebar-\frac{2}{3} N_1 \cdot k\right)\\*
     &+
     \theta(n_3 \cdot k - n_\threebar\cdot k) \theta(n_1 \cdot k - n_\onebar\cdot k)\delta\left(q_\twobar-\frac{2}{3} N_2 \cdot k\right)\\*
     &+
     \theta(n_1 \cdot k - n_\onebar\cdot k) \theta(n_2 \cdot k - n_\twobar \cdot k)\delta\left(q_\threebar-\frac{2}{3} N_3\cdot k\right)
\end{align}

The matrix element for Eikonal emission of one gluon off of 3 Wilson lines is the same as for direct photon production~\cite{Becher:2009th,Schwartz:2016olw} or hard $W/Z$ production~\cite{Becher:2011fc,Becher:2012xr}.
There are 3 Wilson lines in the trijet configuration, pointing in the $n_1,n_2$ and $n_3$ directions (see Fig.~\ref{fig:trijet}). When the jet in the 1 direction is a gluon, the 1-loop soft function is
\begin{multline}
  S_{6g} (q_i) = 2 g_s^2 \mu^{2 \varepsilon} \int \frac{d^d k}{(2 \pi)^{d - 1}}
  \delta (k^2) \theta (k_0) M(k,q_i)\\
  \times \left[ \left( C_F - \frac{1}{2} C_A \right) 
  \frac{n_2\cdot n_3}{(n_2 \cdot k) (n_3 \cdot k)} +
  \frac{1}{2} C_A \frac{n_1 \cdot n_2}{(n_1 \cdot k) (n_2 \cdot k)} + 
  \frac{1}{2} C_A \frac{n_1 \cdot n_3}{(n_1 \cdot k)  (n_3 \cdot k)} \right]
\end{multline}
The soft function with a quark Wilson line in the 1 direction has the color structures interchanged:
\begin{multline}
  S_{6q} (q_i) = 2 g_s^2 \mu^{2 \varepsilon} \int \frac{d^d k}{(2 \pi)^{d - 1}}
  \delta (k^2) \theta (k_0)  M(k,q_i)\\
  \times \left[ \left( C_F - \frac{1}{2} C_A \right) 
  \frac{n_1\cdot n_2}{(n_1 \cdot k) (n_2 \cdot k)} +
  \frac{1}{2} C_A \frac{n_1 \cdot n_3}{(n_1 \cdot k) (n_3 \cdot k)} + 
  \frac{1}{2} C_A \frac{n_2 \cdot n_3}{(n_2 \cdot k)  (n_3 \cdot k)} \right]
\end{multline}

\begin{figure}
    \centering
	\includegraphics[scale=1]{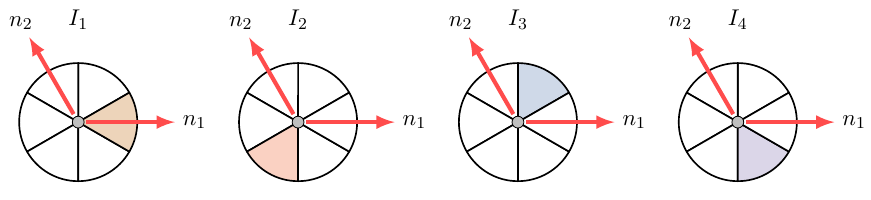}
    \caption{There are 4 independent integrals needed for the soft function. The shaded region indicates the measurement region which can be in various positions relative to the Wilson lines.}
    \label{fig:softregions}
\end{figure}

Despite the preponderance of directions, the integrals required are all of the same general form.
By rotational invariance, we can always take the Wilson lines to be the $n_1$ and $n_2$ directions.
Then all the required integrals are special cases of the general form
\begin{equation}
  I_{n_a,n_b,N} (q) = \int d^d k
  \frac{{n_1} \cdot {n_2}}{({n_1} \cdot
  k) ({n_2} \cdot k)} \delta (k^2) \theta (k^0)
  \delta \left(q- \frac{2}{3}N \cdot k\right) \theta ({n_a} \cdot
  k - {\bar{n}_a} \cdot k) \theta ({n_b} \cdot
  k - {\bar{n}_b} \cdot k) \label{I3form}
\end{equation}
This integral is Lorentz invariant, so it can only depend on dot-products of the 4-vectors involved and is also invariant under separate rescaling of all the $n_i$. Related integrals, with a single $\theta$-function, appear in the iterative solution of the BMS equation~\cite{Banfi:2002hw} for non-global logarithms of the light-jet mass distribution. There, a larger $\text{SL}(2,R)$  symmetry constrains the functional form even more~\cite{Schwartz:2014wha}. Here, the $\text{SL}(2,R)$  is broken by the second $\theta$-function, so the integration region is a cats-eye shaped wedge inside the Poincare disk. However the conformal coordinates proposed in~\cite{Schwartz:2014wha} can still provide a useful change of variables which we used to understand and simplify the integrals. 

In the regions without a Wilson line, the anomalous dimension of the soft function is insensitive to the projection vectors $N_i$; it only depends on the location of the measurement  region relative to the Wilson lines. Thus for NLL resummation there are only 4 independent integrals, as illustrated in Fig.~\ref{fig:softregions}. 
A detailed calculation of the soft integrals can be found in Appendix \ref{appendix:soft}. Here we just summarize the results. 
We find for the 4 integrals

\begin{align}
  I_1(q) &=I_{n_2,n_3,n_1}(q)= \frac{1}{q^{1+2\epsilon}} \left(\frac{1}{\epsilon}
   -\frac{7}{2} \ln 2 + \ln 3 - \frac{3\kappa}{2\pi}  \right)\\
  I_2(q) &= I_{n_1,n_2,n_3}(q)=\frac{1}{q^{1+2\epsilon}} \left( -\ln 2 + \frac{3\kappa}{\pi} \right)\\
  I_3(q) &=I_{\bar{n}_1,\bar{n}_2,N_i}(q)= \frac{1}{q^{1+2\epsilon}} \left(\ln 2 + \frac{3 \kappa}{\pi}\right)\\
  I_4(q) &=I_{\bar{n}_1,\bar{n}_3,N_i}(q)= \frac{1}{q^{1+2\epsilon}} \left(\frac{3}{2} \ln 2 - \frac{3\kappa}{2\pi} \right)
     \label{eq:four_soft_result}
\end{align}
where
\begin{equation}
    \kappa = \text{Im}\, \text{Li}_2\, e^{\frac{\pi i}{3}} \approx 1.0149 \label{kappa}
\end{equation}
is Gieseking's constant. Gieseking's constant is a transecendentality-2 number\footnote{It has not been proven whether Gieseking's constant or Catalan's constant are transecendental, or even irrational. In this context, transecendentality-2 refers to the representation of $\kappa$ as a 2-fold iterated polylogarithic integral.
}  in the family with Catalan's constant $C = \text{Im}\ \text{Li}_2 e^{\frac{\pi i}{2}}$ and  $\pi^2 = 6\text{Li}_2(1)$. 

Then, when we add in the colors structures, the soft function is
\begin{multline}
    I_{6g}(q_i) \propto \delta(q_1)\delta(q_2)\delta(q_2)
    \delta(q_\onebar)\delta(q_\twobar)\delta(q_\threebar) \\*
    + 
    \delta(q_2)\delta(q_2)\delta(q_\onebar)\delta(q_\twobar)\delta(q_\threebar)  \left[\left( C_F - \frac{1}{2} C_A \right)I_2(q_1) +C_A I_1(q_1)\right]
    \\*
    + 
    \delta(q_1)\delta(q_2)\delta(q_3)
    \delta(q_\twobar)\delta(q_\threebar)  \left[\left( C_F - \frac{1}{2} C_A \right)I_3(q_\onebar) +C_A I_4(q_\onebar)\right] + \cdots
\end{multline}
and so on for the other four $q_i$ sectors and for $I_{6q}(q_i)$. For the trijet hemisphere soft function in Eq.~\eqref{trijethemi}, we can set all the $q_i$ in each hemisphere equal For the channel with a gluon jet in the light hemisphere we find
\begin{multline}
    S^\hemi_g(q_\ell,q_h,\mu) \propto \delta(q_\ell)\delta(q_h) \\+ \delta(q_\ell)
    \left[\left( C_F - \frac{1}{2} C_A \right)(I_2(q_h)+2I_3(q_h)) +C_A (2I_1(q_h)+I_4(q_h))\right]\\
     + \delta(q_h)
    \left[\left( C_F - \frac{1}{2} C_A \right)(I_2(q_\ell)+2I_3(q_\ell)) +C_A (2I_1(q_\ell)+I_4(q_\ell))\right]
\\
=\delta(q_\ell)\delta(q_h) + \alphs{\mu}\delta(q_\ell)
    \left[\frac{-4C_F \Gamma_0 \ln \frac{q_h}{\mu} +2\gamma_{sqq} }{q_h}\right]_\star 
    + \alphs{\mu}\delta(q_h)
    \left[\frac{-2C_A \Gamma_0 \ln \frac{q_\ell}{\mu}+ 2 \gamma_{sg}}{q_\ell}\right]_\star 
\end{multline}
with $\Gamma_0=4$ and 
\begin{equation}
    \gamma_{s qq} = -4 C_F \ln 6 ,\quad \gamma_{sg} =-2C_A \ln 3 + 4C_F \ln 2  \label{gammasg}
\end{equation}
Notation for the $\star$ distributions can be found in~\cite{Becher:2006mr,Schwartz:2007ib,Becher:2008cf,Becher:2009th,Becher:2011fc}.

In the  channel where the light hemisphere has quark jet,  the trijet hemisphere soft function has terms of the form
\begin{multline}
    S^\hemi_q(q_\ell,q_h,\mu) \propto \delta(q_\ell)\delta(q_h) \\+ \alphs{\mu}\delta(q_\ell)
    \left[\frac{-2(C_F+C_A) \Gamma_0 \ln \frac{q_h}{\mu} +2\gamma_{sqg} }{q_h}\right]_\star 
    + \alphs{\mu}\delta(q_h)
    \left[\frac{-2C_F \Gamma_0 \ln \frac{q_\ell}{\mu}+ 2\gamma_{sq}}{q_\ell}\right]_\star 
\end{multline}
where 
\begin{equation}
    \gamma_{sqg} = -2(C_A+C_F)\ln 6 ,\quad \gamma_{sq} = - 2 C_F \ln\frac{3}{2} + 2C_A \ln 2  \label{gammasq}
\end{equation}



\subsection{Resummation \label{sec:resum}}
To resum the large Sudakov shoulder logarithms, we convolve the resummed hard, jet and soft function. The resummation of these individual functions is the same as for thrust in the threshold limit~\cite{Fleming:2007xt,Schwartz:2007ib,Becher:2008cf} and other processes~\cite{Becher:2006mr,Becher:2007ty,Becher:2009th,Chien:2010kc,Stewart:2010tn,Jouttenus:2011wh,Feige:2012vc}.

The resummed quark and gluon jet functions have the  form~\cite{Becher:2009th}:
 \begin{equation}
     J_i(m^2,\mu)= 
     \exp\Big[-4C_i S(\mu_j,\mu) + 2 A_{\gamma_{j}} (\mu_j,\mu)\Big]
     \widetilde{j}_i (\partial_{\eta_{j}})
 \frac{1}{m^2} \left( \frac{m^2}{\mu_j^2} \right)^{\eta_{j}} \frac{e^{-\gamma_E \eta_{j}}}{\Gamma(\eta_{j})}
\end{equation}
where the Laplace transform of the 1-loop jet functions are
\begin{equation}
    \tilde{j_i}(L) = 1+ \left(\frac{\alpha_s(\mu_j)}{4\pi}\right)
    \left[C_i \Gamma_0 \frac{L^2}{2}+ \gamma_{i} L\right]
\end{equation}
and the Casimirs and 1-loop anomalous dimensions are
\begin{equation}
     C_q = C_F,\quad C_g = C_A,\quad  \gamma_{jq} = -3C_F,\quad \gamma_{jg} = -\beta_0
     \label{gammaj}
\end{equation}
The Sudakov RG kernel is
\begin{equation}
    S(\nu,\mu) =-
    \int_{\alpha_s(\nu)}^{\alpha_s(\mu)}
    d\alpha \frac{\gamma_\cusp(\alpha)}{\beta(\alpha)}
    \int_{\alpha_s(\nu)}^\alpha \frac{d \alpha'}{\beta(\alpha')}
    = -\frac{\alpha_s}{8\pi} \Gamma_0 \ln^2\frac{\nu}{\mu} + \cdots
\end{equation}
with 
\begin{align}
    \gamma_\cusp(\alpha_s)
    &= \left(\frac{\alpha_s}{4\pi}\right)\Gamma_0 
    + \left(\frac{\alpha_s}{4\pi}\right)^2\Gamma_1 +\cdots\\
    \beta(\alpha_s)
    &=-2\alpha_s \left[\left(\frac{\alpha_s}{4\pi}\right)\beta_0 
    + \left(\frac{\alpha_s}{4\pi}\right)^2\beta_1 +\cdots\right]
\end{align}
where
\begin{align}
\Gamma_0 &= 4,\quad
\Gamma_1 = 4\left[C_A\left(\frac{67}{9}-\frac{\pi^2}{3}\right) - \frac{20}{9} T_F n_f \right] \\
\beta_0 &= \frac{11}{3}C_A - \frac{4}{3} T_F n_f,\quad
\beta_1 = \frac{34}{3} C_A^2 - \frac{20}{3} C_A T_F n_f - 4 C_F T_F n_f
\end{align}
To NLL order
\begin{equation}
    A_{\gamma_j}(\nu,\mu) =  -\gamma_j  \int_{\alpha_s(\nu)}^{\alpha_s(\mu)}
    d\alpha \frac{\alpha}{4\pi\beta(\alpha)}
    =\frac{\gamma_j}{2\beta_0} \ln \frac{\alpha_s(\mu)}{\alpha_s(\nu)}
\end{equation}
Finally,
\begin{equation}
    \eta_{jq} = 2 C_F A_\Gamma(\mu_j,\mu)\ ,\ 
    \eta_{jg} = 2 C_A A_\Gamma(\mu_j,\mu)\ ,
\end{equation}
where
\begin{equation}
    A_\Gamma(\nu,\mu) =  -  \int_{\alpha_s(\nu)}^{\alpha_s(\mu)}
    d\alpha \frac{\gamma_\cusp(\alpha)}{\beta(\alpha)}
        = \frac{\alpha_s}{4\pi} \Gamma_0 \ln \frac{\nu}{\mu} + \cdots
\end{equation}

The hard function can be extracted from~\cite{Ellis:1980wv}, or using the general forms for hard functions in~\cite{Becher:2009qa} or from the hard function for $n$-jettiness~\cite{Jouttenus:2011wh}. It is
\begin{equation}
    H(Q,\mu) = \exp\Big[(4C_F + 2C_A) S(\mu_h,\mu) - 2 A_{\gamma h}(\mu_h,\mu)\Big] \left( \frac{Q^2}{\mu_h^2}\right)^{-(2C_F+C_A)A_\Gamma(\mu_h,\mu)}
    H(Q,\mu_h)
\end{equation}
where
\begin{equation}
    H(Q,\mu_h) = 1+ \left(\frac{\alpha_s}{4\pi}\right)
    \left[-(2C_F+C_A) \frac{\Gamma_0}{4} \ln^2\frac{Q^2}{\mu_h^2}-\gamma_h \ln \frac{Q^2}{\mu_h^2} \right],
\end{equation}
with
\begin{equation}
     \gamma_{h} = -2(2C_F+C_A)\ln 3 - 6C_F - \beta_0 \label{gammah}
\end{equation}

The trijet hemisphere soft functions can be resummed in exactly the same manner as the hemisphere soft function~\cite{Fleming:2007xt,Schwartz:2007ib,Becher:2008cf,Hoang:2008fs,Chien:2010kc}. At NLL level they factorize into the product of soft functions for each hemisphere:
\begin{align}
    S_g^{\hemi}(k_\ell,k_h,\mu) &= S_g(k_\ell,\mu) S_{qq}(k_h,\mu),
    \\*
    S_q^{\hemi}(k_\ell,k_h,\mu) &= S_q(k_\ell,\mu) S_{qg}(k_h,\mu),
\end{align}
The single-variable soft functions all have the same form
\begin{equation}
    S_i(k,\mu) = \exp\Big[2C_i S(\mu_s,\mu) + 2 A_{\gamma_{si}} (\mu_s,\mu)\Big]
    \widetilde{s}_i (\partial_{\eta_{si}})
\frac{1}{k} \left( \frac{k}{\mu_s} \right)^{\eta_{si}} \frac{e^{-\gamma_E \eta_{si}}}{\Gamma(\eta_{si})}\\
\end{equation}
where
\begin{equation}
    \widetilde{s}_i(L) = 1+ \left(\frac{\alpha_s(\mu_s)}{4\pi}\right)\left[-2C_i \Gamma_0 \frac{L^2}{2} + 2\gamma_i L\right]
\end{equation}
and
\begin{equation}
    \eta_{si} =-2 C_i A_\Gamma(\mu_s,\mu) 
\end{equation}
The only difference is the anomalous dimensions. The coefficient of the Sudakov logs are determined by Casimir scaling as the sum of the color factors for each parton in the hemisphere
\begin{equation}
    C_{g} = C_A, \quad C_{qq} = 2C_F, \quad C_{q} = C_F, \quad \text{and} \quad C_{qg} = C_F+C_A
\end{equation}
The anomalous dimensions $\gamma_{sg},\gamma_{sq},\gamma_{sqq}$ and $\gamma_{sqg}$ are in Eqs.~\eqref{gammasg} and \eqref{gammasq}.

Now we just have to put everything together and perform the integrals in Eqs.~\eqref{doublesig}, \eqref{dsdr1} and \eqref{dsdr2}. Since the various functions after resummation are simply powers, e.g. $J(m^2) \sim (m^2)^{\eta_j-1}$, the integrals are all products or convolutions of powers, which can 
be done directly or through Laplace transforms. 

For thrust, with $t=\tau-\frac{1}{3} > 0$, the core measurement function integral following from Eq.~\eqref{tform} is
\begin{equation}
    \int_0^\infty d x  \int_0^\infty dy ~ x^{a-1} y^{b-1} (x+y-t) \theta(x+y-t)=t^{1+a+b} \frac{\Gamma(a)\Gamma(b)}{\Gamma(2+a+b)} \label{tconvo}
\end{equation}
For the left shoulder of heavy jet mass, the integral is similar, but the sign flip in Eq.~\eqref{rform} as compared to Eq.~\eqref{tform} gives an important change.
\begin{equation}
    \int_0^\infty d x  \int_0^\infty dy\ x^{a-1} y^{b-1} (r+y-x) \theta(r+y-x)
    = r^{1+a+b} \frac{\Gamma(a)\Gamma(b)}{\Gamma(2+a+b)}\frac{\sin(\pi a)}{\sin(\pi (a+b))} \label{rconvo}
\end{equation}
For the right shoulder of heavy jet mass we define $s=-r=\rho-\frac{1}{3}>0$. Then the core integral is
\begin{equation}
    \int_0^\infty d x  \int_0^\infty dy\ x^{a-1} y^{b-1} (y-x-s) \theta(y-x-s)
    = s^{1+a+b} \frac{\Gamma(a)\Gamma(b)}{\Gamma(2+a+b)}\frac{\sin(\pi b)}{\sin(\pi (a+b))} \label{sconvo}
\end{equation}
These integrals are all UV and IR divergent, and so analytic continuation has been used to complete them. We discuss the integrals in more detail in Sections~\ref{sec:ngls}.

Putting everything together and applying algebraic simplifications as in~\cite{Becher:2008cf,Becher:2009th}, we find that all 3 observables can be written in terms of the same RG evolution kernel. For the gluon channels
\begin{align}
 \frac{1}{\sigma_1}  \frac{d\sigma_g}{dt} &= \Pi_g(\partial_\etal,\partial_\etah) t \left( \frac{t Q}{\mu_s} \right)^{\etal} \left( \frac{t Q}{\mu_s}
\right)^\etah   \frac{e^{- \gamma_E (\etal + \etah)}}{\Gamma (2 + \etal + \eta_h)} \label{sigmagt}\\
 \frac{1}{\sigma_1}  \frac{d\sigma_g}{dr} &= \Pi_g(\partial_\etal,\partial_\etah)  r \left( \frac{r Q}{\mu_s} \right)^{\etal} \left( \frac{r Q}{\mu_s}
\right)^\etah   \frac{e^{- \gamma_E (\etal + \etah)}}{\Gamma (2 + \etal + \eta_h)} 
\frac{\sin(\pi \etal)}{\sin (\pi (\etal + \etah))} \label{sigmagr} \\
  \frac{1}{\sigma_1} \frac{d\sigma_g}{ds} &= \Pi_g(\partial_\etal,\partial_\etah)  s \left( \frac{s Q}{\mu_s} \right)^{\etal} \left( \frac{s Q}{\mu_s}
\right)^\etah   \frac{e^{- \gamma_E (\etal + \etah)}}{\Gamma (2 + \etal + \eta_h)} \frac{\sin(\pi \etah)}{\sin (\pi (\etal + \etah))} \label{sigmags}
\end{align}
where
\begin{multline}
   \Pi_g(\partial_\etal,\partial_\etah) 
    =  \exp \Big[ 
    4 C_F S (\mu_h, \mu_j )
    + 4 C_F S (\mu_{s}, \mu_j)
    + 2 C_A S (\mu_h, \mu_j) 
    + 2 C_A S (\mu_{s}, \mu_j)\Big]\\
    \times
    \exp \Big[ 
    2 A_{\gamma sg} (\mu_s, \mu_h) 
    + 2 A_{\gamma sqq} (\mu_s, \mu_h) 
    + 2 A_{\gamma j g} (\mu_j, \mu_h) 
    + 4 A_{\gamma j q} (\mu_j, \mu_h) \Big]\\
    \times H (Q, \mu_h)  \widetilde{j}_q \left( \partial_{\eta_h} + \ln \frac{Q \mu_s}{\mu_j^2} \right) \widetilde{j}_{\bar{q}} \left( \partial_{
    \eta_h}+ \ln \frac{Q \mu_{s
}}{\mu_j^2} \right) \widetilde{j}_g \left( \partial_{\eta_\ell} + \ln \frac{Q \mu_{s}}{\mu_j^2} \right) \widetilde{s}_{q q} (\partial_{\eta_h}) \widetilde{s}_g
(\partial_{\eta_\ell}) \label{Pig}
\end{multline}
and
\begin{align}
\etal &= \eta_{jg} + \eta_{sg}  = 2 C_A A_{\Gamma} (\mu_j, \mu_{s})\\
    \etah &= 2 \eta_{j_q} + \eta_{sq q} = 4 C_F A_{\Gamma} (\mu_j, \mu_s)
\end{align}
We have chosen the same jet scales for the light and heavy hemispheres although one could also choose them to be different. Similarly, we have taken the same soft scales for the left and right hemispheres.

One can read off from Eq.~\eqref{sigmagr} that the large logs will be resummed for the left-shoulder of heavy jet mass with the canonical scale choices
\begin{equation}
    \mu_h=Q,\quad \mu_j = \sqrt{r}\, Q,\quad \mu_s = r\, Q \label{canonicalscales}
\end{equation}
For thrust or the right shoulder of heavy jet mass, the canonical scale choices are the same with $r$ replaced by $t$ or $s$ respectively. We have verified that the expansion of the resummed distribution is independent of the matching scales $\mu_h,\mu_j$ and $\mu_s$ at order $\alpha_s$. 

The quark channels have the same form as Eqs.~\eqref{sigmagt} to \eqref{sigmags} but with
\begin{multline}
   \Pi_q(\partial_\etal,\partial_\etah) 
    =  \exp \Big[ 
    (2 C_F +C_A) [S (\mu_h, \mu_j )
    +  S(\mu_{s}, \mu_j)]
    + 2 C_F[ S (\mu_h, \mu_{j}) 
    + S (\mu_{s}, \mu_j)]\Big]\\
    \times
    \exp \Big[ 
    2 A_{\gamma sq} (\mu_s, \mu_h) 
    + 2 A_{\gamma sqg} (\mu_s, \mu_h) 
    + 2 A_{\gamma j q} (\mu_j, \mu_h) 
    + 2 A_{\gamma j q} (\mu_j, \mu_h) 
    + 2 A_{\gamma j q} (\mu_j, \mu_h)\Big]\\
    \times H (Q, \mu_h)  
    \widetilde{j}_q \left( \partial_{\etah} + \ln \frac{Q \mu_s}{\mu_j^2} \right) 
    \widetilde{j}_{g} \left( \partial_{ \etah}+ \ln \frac{Q \mu_s}{\mu_j^2} \right)
    \widetilde{j}_q \left( \partial_{\etal} + \ln \frac{Q \mu_{s}}{\mu_j^2} \right) 
    \widetilde{s}_{q g} (\partial_{\etah}) 
    \widetilde{s}_q(\partial_{\etal})  \label{Piq}
\end{multline}
and
\begin{align}
\etal &= \eta_{jq} + \eta_{sq} = 2 C_F A_{\Gamma} (\mu_j, \mu_{s}) \label{etal}\\
    \etah &= \eta_{jq} + \eta_{jg} + \eta_{sqg}= (2 C_F+2C_A)A_{\Gamma} (\mu_j, \mu_s) \label{etah}
\end{align}
The the final resummed distribution for thrust is
\begin{equation}
    \frac{d\sigma}{dt} = 
    \frac{d\sigma_g}{dt} + 2
    \frac{d\sigma_q}{dt} 
\end{equation}
and similarly for heavy jet mass.


\section{Analysis \label{sec:analysis}}
In Section~\ref{sec:SCET} derived a factorization formula for the left and right Sudakov shoulders for heavy jet mass as well as the right Sudakov shoulder for thrust (thrust has no left shoulder). We will now perform some cross checks on those results. We first perform the fixed order expansion and compare to a numerical computation of the exact NLO expression to verify the singular behavior. Then we demonstrate that there are no non-global logarithms and discuss power corrections.

\subsection{Fixed-order expansions}
First of all, we observe that the full resummed distributions are renormalization-group invariant. This invariance has let us write the evolution kernels in Eqs.~\eqref{Pig} and \eqref{Piq} in a form that depends only on the hard, jet and soft matching scales $\mu_i$, and not on $\mu$. The cancellation of the $\mu$-dependence is non-trivial and requires the Casimirs associated with the Sudakov double logs to cancel and the anomalous dimensions to satisfy
\begin{equation}
    \gamma_h = \gamma_{jg} + 2 \gamma_{jq} + \gamma_{sqq} + \gamma_{sg}
    = \gamma_{jg} + 2 \gamma_{jq} + \gamma_{sqg} + \gamma_{sq} \label{gammacancel}
\end{equation}
These relations can be checked explicitly using Eqs.~\eqref{gammah}, \eqref{gammaj}, \eqref{gammasg} and \eqref{gammasq}.

\begin{figure}
    \centering
    \includegraphics[scale=1]{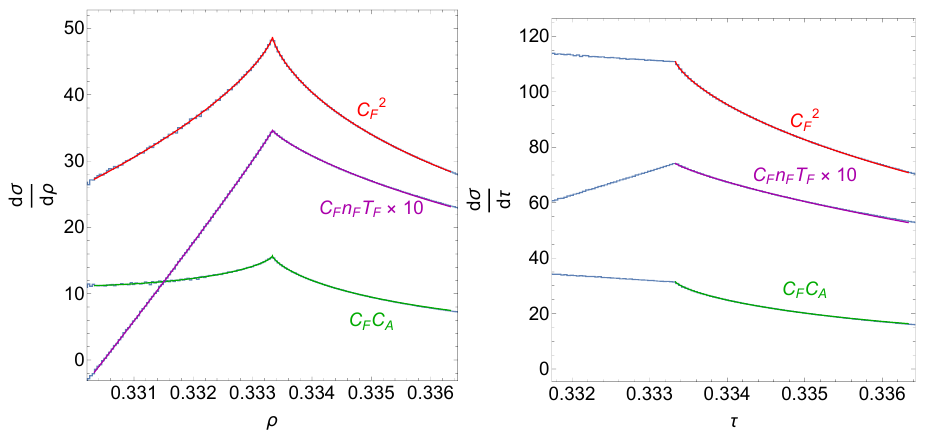}
    \caption{Comparison of the resummed distribution expanded to NLO (colored curves) to the exact NLO distribution in the Sudakov shoulder region (blue histograms) for heavy jet mass (left) and thrust (right). We include in the prediction an offset and a linear term which are fit separately on either side of the peak. The $C_F n_f T_F$ color structure has been scaled up by a factor of 10 for clarity.
The NLO histograms is this figure were computed using {\sc event2} \cite{Catani:1996jh,Catani:1996vz} with a cutoff of $10^{-12}$ and 12 trillion events and normalized to born cross section $\sigma_0$. 
}
    \label{fig:focompare}
\end{figure}

Expanding the resummed distributions to order $\alpha_s$ we find
\begin{multline}
    \frac{1}{\sigma_1}  \frac{d\sigma}{dt} =  3 t 
    +  \frac{\alpha_s}{4\pi} 
    \left\{  B_1 t -\frac{3}{2} (2C_F+C_A)\Gamma_0 t \ln^2 t \right.\\*
    \left.
    \phantom{\frac{3}{2}}
    +\Big[ 3\gamma_{jg} + 6 \gamma_{jq} + 2 \gamma_{sg} + 4 \gamma_{sq} + 2 \gamma_{sqq} +4\gamma_{sqg} +3(C_A+2C_F) \Gamma_0\Big]t \ln t \label{thrustnlogamma}
     \right\}
\end{multline}
for some $B_1$.
The linear terms $3t$ and $B_1 t$ are not predicted with NLL resummation. 
So to be consistent we should remove all the terms linear in $t$. This can be done to all orders by subtracting from the full resummed distribution $\sigma(t)$ the boundary condition $t\,\sigma(1)$. 
That is, we consider
\begin{equation}
     \frac{1}{\sigma_1}  \frac{d\sigma^{\sub}}{dt} \equiv  \frac{1}{\sigma_1}  \frac{d\sigma}{dt} -   t\left[ \frac{1}{\sigma_1}  \frac{d\sigma}{dt}\right]_{t=1} \label{tsubdef}
\end{equation}
which has only terms of the form $t \ln^n t$ to all orders in $\alpha_s$. We use an analogous definition with $t$ replaced by $r$ or $s$ for the subtracted form of the heavy jet mass distribution.
Plugging in the anomalous dimensions 
\begin{multline}
    \frac{1}{\sigma_1}  \frac{d\sigma^{\sub}}{dt} =   \frac{\alpha_s}{4\pi} 
    \left\{
    -6(2C_F+C_A) t \ln^2 t
    \right. \\ \left.
    + \Big[    6C_F(1-4 \ln 3) + C_A (1-12 \ln3)+ 4 n_f T_F  \Big]t \ln t
    \right\} + \cO(\alpha_s^2)
    \label{subexp}
\end{multline}
This is shown in comparison to the NLO calculation in Fig.~\ref{fig:focompare}.

For the left shoulder of heavy jet mass, the expansion gives
\begin{multline}
    \frac{1}{\sigma_1}  \frac{d\sigma^{\sub}}{dr} = \frac{\alpha_s}{4\pi} 
    \left\{
     -\frac{1}{2} (2C_F+C_A)\Gamma_0 r \ln^2 r+
    \Big[ (C_A+2C_F)\Gamma_0 +\gamma_{jg} + 2 \gamma_{jq} + 2 \gamma_{sg} + 4 \gamma_{sq} \Big]r \ln r
    \right\}\label{hjmrgamma}\\*
   =\frac{\alpha_s}{4\pi} 
    \left\{  -2(2C_F+ C_A) r \ln^2 r+\Big[
    2 C_F\left(1+4 \ln \frac{4}{3}\right)
    + C_A \left(\frac{1}{3}+4 \ln\frac{4}{3}\right) +
    \frac{4}{3} n_f T_F \Big] r \ln r
    \right\} 
\end{multline}
This agrees with our fixed-order computation in Section~\ref{sec:FOleft} and with the leading shoulder logarithms at NLO as can be seen in  Fig.~\ref{fig:focompare}. 

Breaking down the expression in Eq.~\eqref{hjmrgamma} the anomalous dimensions which appear are $\gamma_{jg} + 2\gamma_{sg}$ from the gluon channel and $\gamma_{jq} +2 \gamma_{sq}$ from the quark and antiquark channels. So in each channel only anomalous dimensions associated with light-hemisphere side are contributing logarithms as order $\alpha_s$. This is a somewhat remarkable feature of the factorization formula: although both sides contribute 1-loop anomalous dimensions, as is required for renormalization-group invariance, Eq.~\eqref{gammacancel} only one side contributes logarithms. Mechanically, what happens that
\begin{equation}
    (\gamma_h \partial_\etah + \gamma_\ell \partial_\etal) \left[r^{\etal+\etah}-1\right]
     \frac{\sin(\pi \etal)}{\sin(\pi(\etal+\etah))}
    =\gamma_\ell  \, r \ln r 
\end{equation}
So the $\frac{\sin(\pi \etal)}{\sin(\pi(\etal+\etah))}$ factor replaces the full anomalous dimension $\gamma_\ell + \gamma_h$ with just $\gamma_\ell$. 

For the right shoulder of heavy jet mass
\begin{multline}
    \frac{1}{\sigma_1}  \frac{d\sigma^{\sub}}{ds} = \frac{\alpha_s}{4\pi} 
    \left\{
     - (2C_F+C_A)\Gamma_0 s \ln^2 s+
    \Big[2 (C_A+2C_F)\Gamma_0 +2\gamma_{jg} + 4 \gamma_{jq} + 2 \gamma_{sqq} + 4 \gamma_{sqg} \Big] s \ln s
    \right\}\\*
    =\frac{\alpha_s}{4\pi} 
    \left\{  -4(2C_F+ C_A) s \ln^2 s+\Big[
    4 C_F\left(1-4 \ln {6}\right)
    + \frac{2C_A}{3} \left(1-12\ln{6}\right) +
    \frac{8}{3} n_f T_F \Big] s \ln s
    \right\}
\end{multline}
In this case, only the anomalous dimensions in the {\it heavy} hemisphere contribute at NLO. This distribution is also shown in Fig.~\ref{fig:focompare} and compared to the exact NLO calculation. 

\subsection{Non-global logarithms \label{sec:ngls}}
When observables are sensitive to emissions only in a restricted region of phase space, there can be an incomplete cancellation of virtual and real emissions leading to non-global logarithms~\cite{Dasgupta:2001sh}. The classic example is the light-hemisphere mass in $e^+e^-$ collisions. For the light hemsiphere mass, emissions into the heavy hemisphere do not affect the value of the light hemisphere mass, making it non-global. Generally, to be able to resum logarithms using a factorization formula, one would like the condition that the observable be small to force soft and collinear kinematics. This does not happen with the light hemisphere mass, for example, since demanding it be small does not prevent additional hard emissions into the heavy hemisphere. For light-jet mass, the leading non-global logarithm contributes to the cross section at order $d \sigma/d\rho_\ell \sim \alpha_s^2 \ln^2 \rho_\ell$ so it is the same order as terms in NLL resummation. The leading logarithmic series  of non-global logs for the light jet mass and related observables is  understood and can be resummed~\cite{Banfi:2002hw,Schwartz:2014wha}. Progress has also been made on systematic higher-order resummation of non-global logarithms~\cite{Becher:2015hka,Larkoski:2015zka,Banfi:2021owj,Banfi:2021xzn,Becher:2021urs}. Thus, if there were non-global logs in the Sudakov shoulders it would not pose an insurmountable obstacle. Nevertheless, we will show that for the Sudakov shoulders of thrust and heavy jet mass, non-global logs are absent. 

For the right shoulder of thrust, the constraint in Eq.~\eqref{tform} is of the form $t < x+y$ where $x$ and $y$ represent contributions to the mass of the heavy or light hemispheres respectively from soft and collinear radiation near the trijet region (i.e. $x=m_2^2 + m_3^2 + 2 p_2k_2 + 2 p_3 k_3 +2 v_\onebar k_\onebar$ and
$y = m_1^2+ 2 p_1 k_1 + v_\twobar' k_\twobar + v_\threebar'k_\threebar$ when the light jet is in the 1 direction). Since the constraint imposes a {\it lower} bound on $t$, demanding $t\ll 1$ does not force $x$ and $y$ to be small, suggesting that the Sudakov shoulder for thrust might be non-global. However, we can rewrite the core convolution integral in Eq.~\eqref{tconvo} as
\begin{align}
    \int_0^\infty d x  \int_0^\infty dy &~ x^{a-1} y^{b-1} (x+y-t) \Theta(x+y-t) \\*
    &= \int_0^\infty d z \left[\int_0^\infty d x \int_0^\infty  d y\ x^{a-1} y^{b-1} \delta(x+y-z)\right](z-t) \Theta(z-t)\label{square}\\*
    &=\frac{\Gamma(a)\Gamma(b)}{\Gamma(a+b)}\left[
    \underbrace{\int_0^\infty d z\    z^{a+b-1} (z-t)}_{\text{scaleless}}-
    \underbrace{\int_0^t d z\    z^{a+b-1} (z-t)}_{\text{global}}\right]\label{tdrop}
\\*
&=t^{1+a+b} \frac{\Gamma(a)\Gamma(b)}{\Gamma(2+a+b)}
\end{align}
The first integral in brackets in Eq.~\eqref{tdrop} is 
divergent but either independent of $t$ or linear in $t$, so it is smooth across $t=0$ and does not generate Sudakov shoulder logarithms.  The remaining integral has $z=x+y< t$, so taking $t \ll 1$ does force $x,y \ll 1$. We conclude that the right shoulder of thrust should be free of non-global logarithms.  
The actual divergence is an artifact of expanding the phase space limits to leading power. In the full theory, the divergences would cut off by the hard scale $Q$, but still would not generate logarithms of $t$.

It is also worth noting that the scaleless integral in Eq.~\eqref{tdrop} does generate a divergent term proportional to $t$. This would be the same order as terms in the NNLL resummation of the Sudakov shoulder. The presence of such a term does not imply that the factorization formula is valid only to NLL. Indeed, this divergent contribution is smooth across $t=0$, suggesting that it contributes similarly to the left and right sides of $\tau=\frac{1}{3}$ and therefore does not give a discontinuity or a kink at $\tau=\frac{1}{3}$. In any case, since we are only working to NLL in this paper, we can safely ignore it.

For heavy jet mass, the analogous constraint is in Eq.~\eqref{rform} which corresponds to $x < r+y$ for the left shoulder or $x+s<y$ for the right shoulder, as in Eqs.~\eqref{rconvo} and \eqref{sconvo}. We can rewrite Eq.~\eqref{rconvo} as
\begin{equation}
    f(r) = \int_0^\infty d x  \int_0^\infty dy\ x^{a-1} y^{b-1} (r+y-x) \theta(r+y-x)
    = \frac{1}{a(a+1)} \int_0^\infty dy\ (r+y)^{a+1} y^{b-1}  \label{fxint}
\end{equation}
This integral is both UV and IR divergent (for $a,b>0$) and gets contributions from all scales, suggesting, again, that it may generate non-global logarithms. To separate out the UV and IR divergences, we can take two derivatives with respect to $r$, leaving an integral which is UV finite for $a,b>0$. We also introduce a new scale $R$ to separate small $r$ from large $r$. Then we have 
\begin{equation}
    f''(r) 
    = \underbrace{\int_0^R d y\ (r+y)^{a-1} y^{b-1}}_{\text{global}} +\underbrace{\int_R^\infty d y\  (r+y)^{a-1} y^{b-1} }_{\text{regular in $r$}} \label{fpp}
\end{equation}
Since we are interested in the region with $r\ll 1$, we can take $0 < r < R \ll 1$. Then the first integral in Eq.~\eqref{fpp}
is global, since it gets contributions only from the region where $y\ll 1$ and $x\ll 1$ (we had integrated $x$ from 0 to $r+y \ll 1$ in Eq. \eqref{fxint}. The second integral in Eq.~\eqref{fpp} is regular  as $r\to 0$. Thus it does not contribute to any discontinuities or kinks near the shoulder, at $r=0$. As with thrust, it may contribute terms linear in $r$, but will not give any Sudakov shoulder logs. So only the soft and collinear regions should contribute to the Sudakov shoulder logs for heavy jet mass, as with thrust, and there are no non-global logarithms.

To complete the computation, as far as the Sudakov shoulder logs are concerned, we have
\begin{align}
    f''(r) 
    &\cong \int_0^R d y\ (r+y)^{a-1} y^{b-1} \\
    &= \int_0^\infty d y\ (r+y)^{a-1} y^{b-1} -  \underbrace{\int_R^\infty d y\ (r+y)^{a-1} y^{b-1} }_{\text{regular in $r$}}\\
    &\cong  \frac{\Gamma(a) \Gamma(b)}{\Gamma(a+b)} \frac{\sin(\pi a)}{\sin(\pi (a+b))} r^{a+b-1} - R^{1-a-b}\frac{1}{{a+b-1}}
    \label{fpp2}
\end{align}
where we have taken $R \gg r$ to simplify the second integral.
Integrating twice with respect to $r$ then gives
\begin{equation}
    f(r) =\frac{\Gamma(a) \Gamma(b)}{\Gamma(2+ a+b)} \frac{\sin(\pi a)}{\sin(\pi (a+b))} r^{1+a+b}  
    -\frac{r^2}{2} R^{a+b-1}\frac{1}{{1-a-b}}+ c_1 r + c_0
    \label{fcc}
\end{equation}
At small $a$ and $b$ (these are proportional to $\alpha_s$), the second term on the right-hand side is suppressed by a factor of $\frac{r}{R}$ compared to the first term, so it only gives power corrections and no Sudakov shoulder logs, as anticipated.
We should fix the integration constants $c_1$ and $c_2$ so that the expansion of $f(r)$ at small $a$ and $b$ only has terms of the form $r \ln^n r$ with $n>0$. The constant term we can simply discard, $c_0=0$. To fix $c_1$ we should set $c_1 = -f(1)$. This corresponds to integrating $f'(r)$ from $1$ to $r$. These integration constants were used in Eq.~\eqref{tsubdef}.



In summary, the heavy jet mass distribution at a value of $\rho \approx \frac{1}{3}$ does get contributions from phase space regions with jets whose masses are not small. In this sense it is similar to light jet mass near $\rho_\ell = 0$ which gets contributions from phase space regions where $\rho$ is not small. However, the contributions corresponding to heavy jets for the Sudakov shoulder do not generate large logarithms. This is because the phase space regions with heavy jets can contribute to both $\rho \lesssim \frac{1}{3}$ and  $\rho \gtrsim \frac{1}{3}$ and are smooth across $\rho = \frac{1}{3}$. All the contributions to the distribution that are not smooth across $\rho = \frac{1}{3}$ come from the regions with one nearly-massless jet in the light hemisphere and two nearly massless jets in the heavy hemisphere. There is no analog of this continuity argument for light jet mass, which cannot have $\rho_\ell <0$. Thus, the Sudakov shoulders of heavy jet mass (and thrust) are free of non-global logarithms. 

\subsection{Power corrections \label{sec:power}}
In resummed distributions, there are typically different types of power corrections. 
For threshold resummation, near $\rho=0$ for example, there can be power corrections of order $\frac{\LQCD}{Q}$ associated with the strong dynamics of QCD. There can also be hard power corrections,  suppressed by additional powers of  $\rho =\frac{m_H^2}{Q^2}$ where $m_H$ is the mass of the heavy jet. The $\Lambda_\text{QCD}$ power corrections are often modeled with parameters fit to data. This allows for predictivity closer to threshold than with just the resummed distribution alone, although one cannot get too close to threshold since more and more non-perturbative parameters then become relevant.
The hard power corrections are typically accounted for in matching to an exact fixed order expression at large $\rho$.

For Sudakov shoulder resummation, it is not clear whether $\frac{\LQCD}{Q}$ power corrections are important near the trijet threshold. On the one hand, the resummed distribution involves evaluating $\alpha_s$ at scales such as $\mu_s = Q r$ which can reach $\LQCD$ for small enough $r$. On the other hand, the shoulder is intrinsically perturbative, associated with fixed-order phase space boundaries, so one might expect that it might be invisible to non-perturbative physics.

\begin{figure}
    \centering
    \includegraphics[scale=0.7]{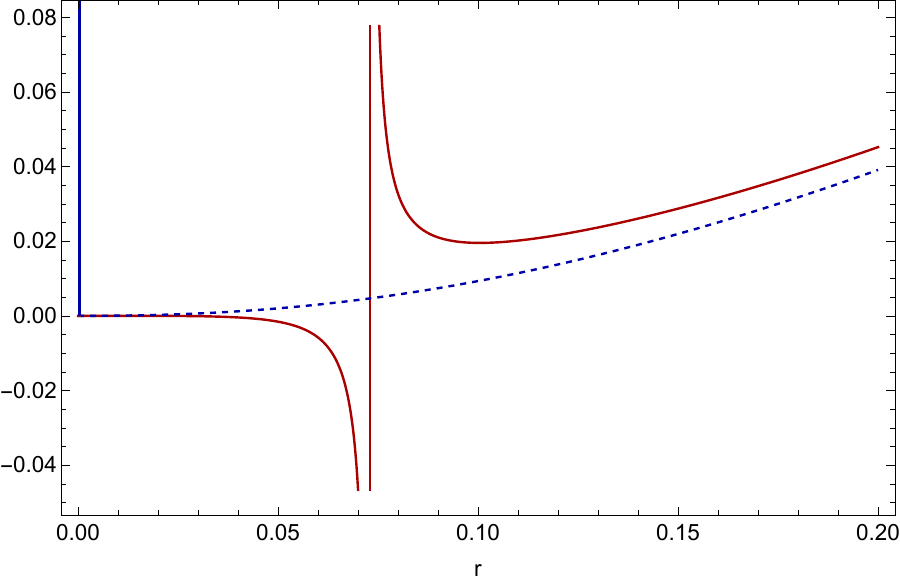}
    \caption{The solid red curve shows $r^{1+a+b}
    \frac{\Gamma(a) \Gamma(b)}{\Gamma(2+a+b)} 
    \frac{\sin(\pi a)}{\sin(\pi(a+b))}$ with $a=b=-6\frac{\alpha_s}{\pi} \ln r$ and $\alpha_s =  0.1$. The pole is at $a+b=1$. The dashed curve shows this same function once the power-suppressed term $-\frac{r^2}{2} R^{a+b-1} \frac{1}{1-a-b}$ is added with $R=1$. The power suppression is apparent at small $r$, where the difference between the two curves is negligible. The pole at $a+b=2$ is not canceled and appears as the spike at $r\approx 0.0003$.
    }
    \label{fig:pole}
\end{figure}

The hard power corrections for the Sudakov shoulder are more interesting. In Section~\ref{sec:ngls} we argued that at leading power all the non-analytic behavior near the shoulder is determined by soft and collinear physics. That is, there are no non-global logarithms. One can see this from Eq.~\eqref{fcc}. The quantities $a$ and $b$ are to be replaced by $\etal$ and $\etah$ in the resummed distribution, which are parameterically of the form $\eta \sim \alpha_s \Gamma_0  \ln r$. Thus at small $\alpha_s$, all the terms of the form $r \ln^n r$ will come from the expansion of the first term on the right-hand side in  Eq.~\eqref{fcc}. On the other hand if $r$ is sufficiently small then $a+b$ can be of order $1$. As $a+b$ nears $1$ a pole from the $\sin^{-1}(\pi(a+b))$ factor in Eq.~\eqref{fcc} is approached. However, when $a+b\approx 1$, power-suppressed term is no-longer power suppressed. Indeed, it has precisely the behavior needed to remove the singular behavior from the leading power term. We show this in Fig.~\ref{fig:pole}.

To see what is happening analytically, noting that the leading power expression scales like $r^{1+a+b}$ we can write the subleading power expression as
\begin{equation}
    r^2 R^{a+b-1}  = r^{1+a+b} \left(\frac{r}{R}\right)^{1-a-b}
\end{equation}
For $a+ b\ll1$ there is a $\frac{r}{R}$ linear power suppression. However for $a+b\sim 1$ there is no power suppression at all; for $a+b=1$ this expressions reduces to $r^2$ as does $r^{1+a+b}$. In effect, the scaling dimensions of the leading power and subleading power pick up such large anomalous contributions that their relative scaling changes.

Taking $R\to \infty$ gives the leading contribution. The first subleading power contribution in this limit cancels the pole at $a+b=1$. To cancel subsequent poles, one can use the exact integrated form of Eq.~\eqref{fpp}. This effectively replaces $f(r)$ in Eq.~\eqref{fcc} by
\begin{equation}
    f(r) =r^{1+a+b}e^{-i b \pi} \frac{1}{a(a+1)} B_{-\frac{R}{r}}(b,2+a)+ c_1 r + c_0
\end{equation}
where $B_z(x,y)$ is the incomplete Euler $\beta$ function and $c_0$ and $c_1$ are again integration constants to be fixed with physical boundary conditions.


\section{Discussion \label{sec:discussion}}
Next, we want to evaluate the resummed distribution numerically and compare to fixed order, to see the effect of the higher order logarithms. There are a number of issues which complicate the analysis, compared with typical threshold resummation of large logarithms. 

First, the relevant domain of the observable is rather small for Sudakov shoulders. For example, for thrust in the threshold limit, although the logarithms are largest at small $\tau$, power corrections and subleading logs are also large there. Typical fits restrict $\tau \gtrsim 0.1$ where perturbative control is best. For example, with $Q=92$ GeV, Ref.~\cite{Becher:2008cf} used $0.1 < \tau < 0.24$ for their $\alpha_s$ fits to thrust while Ref.~\cite{Abbate:2010xh} took $\tau \gtrsim \frac{6\,\text{GeV}}{Q} = 0.066$. For the right shoulder of thrust which begins at the 3-parton maximum $\tau=\frac{1}{3}$ if one excludes the region up to $\frac{1}{3}+ 0.1 = 0.43$ there is no cross section or phase space left! Moreover, the 4-particle phase space forces $\tau \lesssim 0.42$, so there is another Sudakov shoulder at this thrust value whose logs must be resummed separately. So it is not clear if there is a region on the right shoulder where the resummed formula might even be valid. For the left shoulder, in contrast, one can exclude the region with $r = \frac{1}{3} - \rho < 0.1$ which still leaves a region of $0.1 \lesssim \rho \lesssim 0.23$ in which Sudakov shoulder logarithms might be important and renormalization-group improved perturbation theory could be valid.

Second, in the threshold region, the logarithms of thrust are of the form $\frac{d\sigma}{ d \tau } \sim \alpha_s^n \frac{\ln^m \tau}{\tau}$. In contrast, the logarithms near the shoulder region are of the form $\frac{d\sigma}{ d t } \sim \alpha_s^n t \ln^m t$. So they are suppressed effectively by $t^2$ compared to the threshold region. The thrust and heavy mass distributions are indeed finite at the trijet threshold to all orders while they are divergent at the dijet threshold. Despite this additional suppression, the logarithms are noticeable, as can be been in Fig.~\ref{fig:thrustandhjm}.

Third, in the important left-shoulder region for heavy jet mass, the resummed distribution has usually singular behavior. Let us recall the form of the resummed heavy jet mass distribution in the region $r=\frac{1}{3} - \rho \ll 1$ when the light hemisphere has a gluon jet from Eq.~\eqref{sigmagr}
\begin{equation}
   \frac{1}{\sigma_1}  \frac{d\sigma_g}{dr}  = \Pi_g(\partial_\etal,\partial_\etah)  r \left( \frac{r Q}{\mu_{s \ell}} \right)^{\etal} \left( \frac{r Q}{\mu_{s h}}
\right)^\etah   \frac{e^{- \gamma_E (\etal + \etah)}}{\Gamma (2 + \etal + \eta_h)} 
\frac{\sin(\pi \etal)}{\sin (\pi (\etal + \etah))}  \label{Rg}
\end{equation}
with $\etah$ and $\etal$ in Eqs.\eqref{etah} and \eqref{etal}. This expression has singularities whenever $\etal + \etah \in \mathbb{Z}$.

Choosing canonical scales as in Eq.~\eqref{canonicalscales} at leading logarithmic level gives
\begin{equation}
    \etal + \etah  =-\frac{\alpha_s}{2\pi} (C_A+ 2C_F) \Gamma_0 \ln \frac{\mu_j}{\mu_s} = \frac{\alpha_s}{4\pi} (C_A+ 2C_F) \Gamma_0 \ln r  \label{LLeta}
\end{equation}
The singularity $\etal + \etah = 0$ occurs when $\mu_j=\mu_s$, which happens at $r=1$. At $r=1$ there are no logarithms, so this singularity is entirely removed by the subtraction in Eq.~\eqref{tsubdef}. 
That is,
\begin{equation}
\frac{d\sigma^\text{sub}}{dr} 
=
\frac{d\sigma}{dr} -
r \left[
\frac{d\sigma}{dr} \right]_{r=1}
\end{equation}
is regular at $r=1$. 
Note, however, if the soft and jet scales meet at some lower scale, this singularity may be reintroduced.

The singularity at $\etal + \etah = 1$ is more troublesome. Similar singularities have been seen in other processes, such as Drell-Yan or Higgs production at small $p_T$~\cite{Catani:1996yz,Frixione:1998dw,Becher:2010tm,Monni:2016ktx,Banfi:2018mcq} 
or the jet shape~\cite{Seymour:1997kj,Cal:2019hjc}.
Writing $L=\ln \frac{1}{r}$,
resummation at order NLL is meant to get right all terms of order $\alpha_s^n L^j$ with $j\ge 2n-1$ in $R(r)$ or equivalently all terms of order $\alpha_s^n L^j$ with $j\ge n$ in the exponent, i.e. in $\ln R(r)$.
In the notation of~\cite{Catani:1992ua}, we can write
\begin{equation}
    \ln
    \frac{d\sigma^\text{sub}}{dr}
    =  L g_1 (\alpha_s L) + g_2 (\alpha_s L) + \cdots =  \sim \cdots - \ln \sin(\pi(\etal + \etah)) + \cdots
\end{equation}
with $g_1(\alpha_s L)$ and $g_2(\alpha_s L)$ completely fixed by the expansion and reorganization of our resummed expression.  Normally, when $\alpha_s L \sim 1$ then we must go to higher order in RG-improved perturbation theory; at NNLL level, we would have additionally $L g_3(\alpha_s L)$ which would extend the validity of the theoretical prediction. Here, instead we find a singularity in the exponent: $ \ln R^\text{sub}$ is infinite at $\alpha_s L \sim 1$ due to the $\etal + \etah = 1$ singularity. Therefore, going beyond NLL would {\it not} allow us to make perturbative predictions beyond where the singularity occurs. Instead, the singularity is canceled by including subleading power effects, as discussed in Section~\ref{sec:power} and shown in Fig.~\ref{fig:pole}.

The singularity at $\alpha \ln r \sim 1$ is reminiscent of the Landau pole in QCD. There, already at 1-loop one can see a pole in the running coupling at $\mu=\Lambda_\text{QCD}$. With 2-loop or higher order running, the precise location of $\Lambda_\text{QCD}$ moves around, but cannot be surpassed. Thus what we see here is a kind of Sudakov Landau pole. Using the LL form in Eq.~\eqref{LLeta} it occurs at
\begin{equation}
    r = \exp\left[-\frac{4\pi}{(C_A+2C_F)\alpha_s \Gamma_0}\right] = \exp\left[ -\frac{3\pi}{17\alpha_s}\right]
\end{equation}
For $\alpha_s = 0.119$ this gives $r\approx 0.01$.
Using the NLL expressions for $\etal$ and $\etah$ with canonical scale choices, Eq.~\eqref{canonicalscales} the pole ascends to $r \approx 0.06$. Thus we cannot expect the leading-power NLL resummed distribution to be predictive between $0.27 < \rho < 0.39$. This essentially excludes the entire region on the right shoulder, but leaves the region with $\rho \lesssim 0.27$ as potentially viable for a precise prediction. To stay well away from the singular region however, one must take $\rho$ smaller, $\rho \lesssim 0.2$ where the logarithms are no longer particularly large.

We emphasize that excluded range is larger than that associated with strong coupling. With 2-loop running and $\alpha_s(m_Z) = 0.119$, we find $\alpha_s(m_s) = 1$ at $r=0.005$. Thus the singularity comes in at a factor of 10 larger values of $r$ than where the soft scale probes strong dynamics. This is because the singularity is associated with the cusp anomalous dimension, not the QCD $\beta$ function: the two Landau poles are unrelated. 

Because of the Sudakov Landau pole  in the resummed distribution it is difficult to make quantitative predictions, particularly at the NLL level, without a better understanding of the power corrections. 
There are a number of approaches that could be applied to ameliorate the problem. In~\cite{Frixione:1998dw}, a similar pole in the Drell-Yan spectrum at small $p_T$ 
(at $q^\star =m_Z \exp(-\frac{3\pi}{8\alpha_s})$~\cite{Becher:2011xn})
was shown to be associated with power-suppressed region of small impact parameter, but could be softened with higher-order resummation.
In~\cite{Monni:2016ktx} it is argued that one could also do resummation in momentum space directly with a modified expansion of the Sudakov radiator. Related ideas can be found in~\cite{Ebert:2016gcn,Cal:2019hjc}.
It will be important to understand which of these approaches might apply for Sudakov shoulder resummation, but we do not attempt a complete analysis here.

\begin{figure}
    \centering
    \includegraphics[scale=0.7]{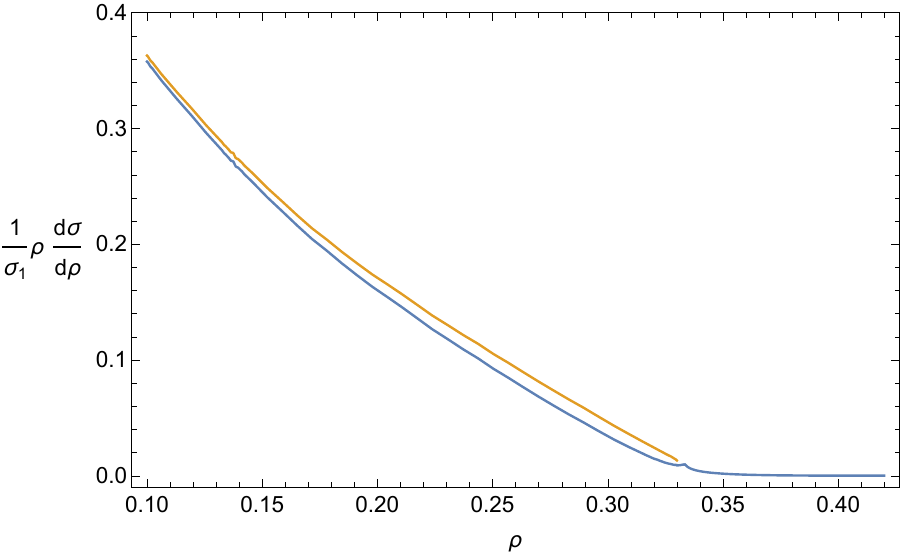}
    \caption{Resummation of the left Sudakov shoulder for heavy jet mass at leading-logarithmic level (upper curve) compared to NLO (lower curve). The strong coupling constant is fixed to $\alpha_s = 0.119$}
    \label{fig:LL}
\end{figure}

At the LL level, however, because the Sudakov Landau pole is very close to the shoulder, we can at least begin to get a quantitative feel of how important resummation is.
Consider the LL distribution using canonical scales in Eq.~\eqref{canonicalscales}. When the jet in the light hemisphere is a gluon, it has the form as in Eq.~\eqref{Rg} with $\Pi_g$ from Eq.~\eqref{Pig} becoming
\begin{equation}
    \Pi_g = e^{-\frac{\alpha_s}{8\pi} \Gamma_0 (C_A+2C_F) \log ^2r}
    \left[1-\frac{\alpha_s}{8\pi}\Gamma_0
    \left(C_A \partial^2_\etal + 2C_F \partial_\etah^2\right)\right]
\end{equation}
Note that we include every term with $\Gamma_0$ in it for leading-log resummation, not just the exponential prefactor. Including only the prefactor would  give the double-logarithmic approximation, as used in previous work on resummation of the $C$ parameter Sudakov shoulder~\cite{Catani:1997xc}. We subtract off from the resummed distribution $r$ times its $r\to 1$ limit as done in Eq.~\eqref{tsubdef}. Note that this subtraction must be done before setting canonical scales. We then match to the fixed order LO+NLO calculation by subtracting from the resummed distribution its expansion to order $\alpha_s$. In this case, the matching subtraction is
\begin{equation}
   \frac{1}{\sigma_1}  \frac{d\sigma^\text{match}}{dr} = \frac{\alpha_s}{4\pi}(C_A+2C_F)  \Gamma_0 (  r\ln r - \frac{1}{2} r\ln^2 r)
\end{equation}
Finally we include the subtraction of the first subleading power contribution. For the gluon channel this amounts to subtracting:
\begin{equation}
 \frac{1}{\sigma_1}  \frac{d\sigma_g^R}{dr} =
 \frac{r^2}{2R}\Pi_g(\partial_\etal,\partial_\etah)  
  \left( \frac{R Q}{\mu_{s \ell}} \right)^{\etal} 
 \left( \frac{R Q}{\mu_{s h}}
\right)^\etah  
\frac{e^{- \gamma_E (\etal + \etah)}}
{\Gamma(\etal)\Gamma(\etah)}\frac{1}{ (1-  \etal - \eta_h)}
\end{equation}
from the resummed distribution. 
The resulting LL resummed and matched result at $Q=m_Z$ and $\alpha_s(Q)=0.119$ with $R=\frac{1}{3}$ for the left shoulder is shown in Fig.~\ref{fig:LL}. To be clear, this is not the complete power correction, but amounts to integrating the leading power soft and collinear matrix elements outside of their formal region of validity upto the kinematic limit of $r=\frac{1}{3}$.

\section{Conclusion \label{sec:conc}}
Thrust $\tau$ and heavy jet mass $\rho$ are two of the most important observables at $e^+e^-$ colliders. They have been used for decades for tests of precision QCD and measurements of $\alpha_s$. 
At leading order in perturbation theory, both $\rho$ and $\tau$ are phase-space limited to be less than $\frac{1}{3}$ and have a non-vanishing slope as $\frac{1}{3}$ is approached. At next-to-leading order, thrust behaves like $\alpha_s^2 (\tau-\frac{1}{3}) \ln^2 (\tau-\frac{1}{3})$ for $\tau>\frac{1}{3}$ so that the slope diverges as $\frac{1}{3}$ is approached from the right. This behavior is called a right Sudakov shoulder. Heavy jet mass has a slope which diverges as $\rho$ nears $\frac{1}{3}$ both from the left and the right: it has two Sudakov shoulders. The left shoulder of heavy jet mass is particularly important as the large logarithms can extend well into the region where $\alpha_s$ fits are typically done ($0.1 \lesssim \rho \lesssim 0.24$). Thus understanding and resumming its Sudakov shoulders could be very important for improving agreement of theoretical predictions with data and subsequent extractions of $\alpha_s$. We also point out that it has been noted recently in the literature that in the context of other event shape observables such as fractional moments of energy-energy correlation \cite{Banfi:2018mcq} or projected energy correlators \cite{Chen:2020vvp}, one must resort to a joint resummation of the Sudakov shoulders and endpoint peaks.

We derived a factorization formula for both thrust and heavy jet mass in the Sudakov shoulder region. The basic mechanism for generating Sudakov shoulder logs is when a soft or collinear emission goes into one hemisphere a global constraint such as $m^2 < \rho - \frac{1}{3}$ transfers large logs from the emissions to the shoulder. Although the constraint seems non-local, involving both hemispheres, and therefore might violate factorization, we show that it does not. Moreover regions of large jet mass do not contribute Sudakov shoulder logs, showing that there is no non-global log contribution in the shoulder region. We checked our factorization formula by expanding to NLO and comparing to the exact numerical NLO calculation very close to the shoulder region. As can be seen in Fig.~\ref{fig:focompare} the agreement is excellent. 

The calculation involves some unusual ingredients. Since the emissions come off a trijet configuration with 2 quarks and 1 gluons, there is no azimuthal symmetry (unlike the threshold case), and the polarization of the gluon affects the spectrum. At leading order, only a uniform azimuthal angle integral was needed for the resummed expression, but in general polarized splitting function may be necessary. We also saw the appearance of Gieseking's constant, a transcendentally-two number. Although it also drops out of the NLL expression, at higher orders it or related constants may be involved.

The resummed distribution for heavy jet mass has a term of the form $\sin^{-1} (\pi \eta)$ with $\eta \sim \alpha_s \Gamma_0 \ln r$, where $r=\frac{1}{3} - \rho$. The expansion near $\alpha_s = 0$ (or $\ln r = 0$) produces the leading and next-to-leading logarithmic series: terms like $\alpha^n \ln^{2n} r$. However, there is also a pole at $\eta = 1$. This pole in the resummed distribution is not due to the running coupling -- it is present even with $\beta(\alpha_s) = 0$ -- but due to the cusp anomalous dimension. Thus it is a kind of Sudakov Landau pole. Similar behavior has been seen before, in the Drell-Yan process at small $p_T$, for example~\cite{Frixione:1998dw,Catani:1996yz,Monni:2016ktx}. In both cases there is a connection between the pole and subleading power effects (subleading in $r$ for the shoulder, or in impact parameter $b$ for Drell-Yan). We show that subleading power terms can in fact cancel the $\eta=1$ pole but do not affect the NLL series. This implies that a better understanding of power corrections will be necessary to establish proper theoretical uncertainty on the resummed distribution. There are many approaches that may help improve the convergences of the resummed distribution~\cite{Becher:2007ty,Ebert:2016gcn,Monni:2016ktx}.

Although our results are only valid to NLL level, the factorization formula applies to all orders. In fact, since the anomalous dimensions of the jet and hard functions are known to two loops, and therefore the soft function anomalous dimension as well by renormalization-group invariance, NNLL resummation should be possible. At NNLL level, terms linear in $\rho$ or $\tau$ are determined. The slope can be discontinuous from the left to right side of the shoulder, as it is already at LO. This discontinuity should be computable. However, because there is also a linear term in the distribution not associated with the shoulder, confirming the predictions at NNLL will be challenging. Nevertheless, pushing the limits of Sudakov shoulder resummation, not just for $e^+e^-$ event shapes but for collider observables more broadly, provides opportunities to improve our understanding of precision QCD.

\section*{Acknowledgements}
The authors would like to thank Vicent Mateu for valuable conversations that helped inspire this work, as well as Thomas Becher, Stafano Catani, Pier Monni and Bryan Webber for valuable feedback. This study was supported by the U.S. Department of Energy under contract DE-SC0013607.

\newpage
\appendix
\section{Calculation of the one-loop trijet soft function}\label{appendix:soft}
Before using rotational invariance all the integrals needed for the  one-loop trijet soft functions are of the form 
\begin{equation}
 I_{n_a,n_b,n_c,n_d,N} (q) = \int d^d k
  \frac{{n_a} \cdot {n_b}}{({n_a} \cdot
  k) ({n_b} \cdot k)} \delta (k^2) \theta (k^0)
  \delta \left(q - \frac{2}{3}N \cdot k\right) \theta ({n_c} \cdot
  k - {\bar{n}_c} \cdot k) \theta ({n_d} \cdot
  k - {\bar{n}_d} \cdot k)
\end{equation}
where the $N$ is selected from the direction vectors
\begin{align}
    n_1&=(1,0,0,1),\quad  n_2=\left(1,0,\frac{\sqrt{3}}{2},-\frac{1}{2}\right),\quad n_3=\left(1,0,-\frac{\sqrt{3}}{2},-\frac{1}{2}\right) \label{nas}\\
    N_1&=(2,0,0,-2), \quad N_2=(2,0,\sqrt{3},3), \quad N_3=(2,0,-\sqrt{3},3)
\end{align}
and $n_a\cdots n_d$ only from the $n_j$ in Eq.~\eqref{nas}.
The $\theta$-functions restrict the phase space to one of the sextants in Fig.~\ref{fig:oranges}.

The Wilson lines in the trijet configuration have an $S_3$ symmetry which includes a
$Z_3$ rotational invariance and a reflection symmetry. We can use the rotational invariance to rotate the Wilson lines so that they always point in the $n_1$ and $n_2$ directions. Thus we only need to consider integrals as in Eq.~\eqref{I3form}:
\begin{equation}
  I_{n_a,n_b,N} (q) = \int d^d k
  \frac{{n_1} \cdot {n_2}}{({n_1} \cdot
  k) ({n_2} \cdot k)} \delta (k^2) \theta (k^0)
  \delta \left(q- \frac{2}{3}N \cdot k\right) \theta ({n_a} \cdot
  k - {\bar{n}_a} \cdot k) \theta ({n_b} \cdot
  k - {\bar{n}_b} \cdot k) 
\end{equation}
When $N=n_i$ is one of the Wilson line directions, then only two sextants are relevant, $I_1(q)$ and $I_2(q)$ from Fig.~\ref{fig:softregions}:
\begin{align}
\label{I12def}
    I_1(q)=I_{n_2, n_3, n_1}(q), \quad
    I_2(q)=I_{n_1, n_2, n_3}(q)
\end{align}
When $N=N_1 = 2 \bar{n}_1$, there are two configurations relevant, with both Wilson lines adjacent to the $n_\onebar$ measurement region or just one of them adjacent two it. The two integrals are, as in  Fig.~\ref{fig:softregions},
\begin{align}
\label{I34def}
    I_3(q)=I_{\bar{n}_1, \bar{n}_2, 2 \bar{n}_3}(q),\quad
    I_4(q)=I_{\bar{n}_1, \bar{n}_3, 2 \bar{n}_2}(q)
\end{align}
where $2 \bar{n}_3$ comes from rotating $N_1$ as the Wilson lines are rotated to the $n_1$ and $n_2$ directions.

The remaining integrals involve $N_2$ and $N_3$. These vectors are not lightlike, but they are related by a $Z_2$ symmetry. (Recall that the origin of the asymmetry between $N_1$ and $N_2/N_3$ is that $n_1$ points to the light-hemisphere which affects the soft projections in the factorization formula.) Since $N_2$ and $N_3$ are related by a reflection in the $y$ direction, which is a symmetry of the Wilson lines, if we know the integral for all Wilson line configurations for $N_2$ we know it for $N_3$ as well. So there are 3 possibilities, corresponding to the location of the 3 measurement regions with respect to the Wilson line. Rotating the $N_2$ Wilson line by $\frac{2\pi}{3}$ and $\frac{4\pi}{3}$ gives 
\begin{equation}
    N_2'=(2,0,2\sqrt{3},0), \quad N_2''=(2,0,-\sqrt{3},-3)
\end{equation}
Thus the last 3 integrals we need are
\begin{equation}
    I_5(q)=I_{\bar{n}_1,\bar{n}_2,N_3}(q),\quad
    I_6(q)=I_{\bar{n}_1,\bar{n}_3,N_3'}(q), \quad
    I_7(q)=I_{\bar{n}_2,\bar{n}_3,N_3''}(q)
    \label{I567def}
\end{equation}


To perform the integrals, we parameterize the phase space with lightcone components in some direction $n_1$:
\begin{equation}
k^\mu=k_{+}\frac{n_1^\mu}{2}+k_{-}\frac{\bar{n}_1^\mu}{2}+k_{\perp}^\mu  
\end{equation}
so that
\begin{equation}
    d^d k=\frac{1}{2} d\Omega_{d-2} k_{\perp}^{d-3}d k_{\perp} dk_{+} dk_{-}=\frac{1}{2}d\Omega_{d-3} \sin^{d-4}\theta d\theta k_{\perp}^{d-3}d k_{\perp} dk_{+} dk_{-}
\end{equation}
and
\begin{equation}
     \delta(k^2) = \delta(\vec{k}_{\perp}^2-k_{+}k_{-})
\end{equation}
The integral over $k_{\perp}$ can be calculated using the $\delta$ function and then $k_{+}$ and $k_{-}$ rescaled by $q$ to obtain the $q$ dependence $q^{d-5}$ as expected by dimensional analysis. 
We also introduce $t=\sqrt{\frac{1-\cos\theta}{1+\cos\theta}}$ to rationalize $\sin \theta = \sqrt{1-\cos^2\theta}$.

With these preliminariles, the integral $I_1(q)=I_{n_2, n_3, n_1}(q)$ takes the form in $d=4-2\epsilon$ dimensions
\begin{align}
    I_1&=\frac{3^{1-2\epsilon} \Omega_{1-2\epsilon}}{q^{1+2\epsilon}} \int_0^\infty \frac{dt}{t^{2\epsilon} (1+t^2)^{1-2\epsilon}} \int_0^\infty \frac{dk_{+}}{k_{+}^\epsilon} \frac{1}{1+3k_{+}-\frac{2\sqrt{3k_{+}}(1-t^2)}{1+t^2}} \\
    &\times\theta\left(1+k_{+}\right)\theta\left(-1+k_{+}-\frac{2\sqrt{3 k_{+}}(1-t^2)}{1+t^2}\right) \theta\left(-1+k_{+}+\frac{2\sqrt{3k_{+}}(1-t^2)}{1+t^2}\right)
\end{align}
The $\theta$-functions impose that
\begin{equation}
    0<t<\infty,\quad k_{+}>\frac{7-10t^2+7t^4}{(1+t^2)^2}+\frac{4\sqrt{3(1-t^2)^2(1-t^2+t^4)}}{(1+t^2)^2}\equiv k_c
\end{equation}
To handle the UV divergence as $k_{+}\to \infty$, we can add and subtract the integral $I_2^\text{div}$ over the integrand expanded at large $k^+$. This subtraction term requires the integral
\begin{equation}
     \int_0^\infty \frac{dt}{t^{2\epsilon} (1+t^2)^{1-2\epsilon}}\int_{k_c}^\infty \frac{dk_{+}}{k_{+}^\epsilon}\frac{1}{3k_{+}}=\frac{\pi}{6}\frac{1}{\epsilon}+\frac{1}{9} \left(-5\kappa +\pi\ln\frac{64}{3}\right)+\mathcal{O}(\epsilon)
\end{equation}
where $\kappa= \text{Im}\, \text{Li}_2\, e^{\frac{\pi i}{3}}$ is Gieseking's constant.
Then $I_1^\text{fin} = I_1 - I^\text{div}_1$ is finite and can be expanded in $\epsilon$ and integrated order by order. 
Eventually, we arrive at 
\begin{equation}
    I_1(q)=\pi^{-\epsilon}e^{-\gamma_E \epsilon} \left(\frac{2}{3}\right)^{2\epsilon}\frac{1}{q^{1+2\epsilon}}\left(\frac{1}{\epsilon}+\ln 3-\frac{7}{2}\ln 2-\frac{3}{2\pi}\kappa+\mathcal{O}(\epsilon)\right)
    \label{eq:I2_result}
\end{equation}

The calculation for other soft integrals are similar, and the results up to order $\mathcal{O}(\epsilon)$ are summarized as followed:
\begin{align}    
\label{I1form}
    I_1(q)&=\mathcal{N}\bigg[\frac{1}{\epsilon}+\ln 3-\frac{7}{2}\ln 2-\frac{3}{2\pi}\kappa+\epsilon\bigg(\frac{18}{5\pi}c_1+\frac{8}{\pi}c_2
    -\frac{103}{180}\pi^2+\frac{3}{\pi}\kappa\ln 2+\frac{10}{3}\ln^2 2\notag\\
&-\frac{7}{2}\ln 2\ln 3+\frac{17}{40} \ln^2 3+\frac{5}{12}\text{Li}_2 \left(\frac{1}{4}\right)\bigg)+\mathcal{O}(\epsilon^2)\bigg]\\
    I_2(q)&=\mathcal{N}\bigg[{ \frac{3}{\pi}\kappa-\ln 2}+\epsilon\left(2\ln^2 2+\frac{3}{2\pi}c_3-\frac{6}{\pi}\kappa\ln 2\right)+\mathcal{O}(\epsilon^2)\bigg]\\
    I_3(q)&=\mathcal{N}\left[{ \frac{3}{\pi}\kappa+\ln2}+\epsilon\left(-2\ln^2 2 -\frac{6}{\pi}\kappa\ln 2+\frac{3}{2\pi}c_5\right)+\mathcal{O}(\epsilon^2)\right]\\
    I_4(q)&=\mathcal{N}\bigg[{ -\frac{3}{2\pi}\kappa+\frac{3}{2}\ln 2}+\epsilon\left(-3\ln^2 2+\frac{3}{2\pi}c_4+\frac{3}{\pi}\kappa \ln 2\right)+\mathcal{O}(\epsilon^2)\bigg]\\
    I_5(q)&=\mathcal{N}\bigg[{\frac{3}{\pi}\kappa+\ln 2}+\epsilon\left(-2\ln^2 2+\frac{3}{2\pi}c_7-\frac{6}{\pi}\kappa\ln 2\right)+\mathcal{O}(\epsilon^2)\bigg]\\
    I_6(q)&=\mathcal{N}\bigg[{ -\frac{3}{2\pi}\kappa+\frac{3}{2}\ln 2}+\epsilon\left(-3\ln^2 2+\frac{3}{2\pi}c_6+\frac{3}{\pi}\kappa\ln 2\right)+\mathcal{O}(\epsilon^2)\bigg]\\
    I_7(q)&=\mathcal{N}\bigg[{ -\frac{3}{2\pi}\kappa+\frac{3}{2}\ln 2}+\epsilon\left(-3\ln^2 2+\frac{3}{2\pi}c_8+\frac{3}{\pi}\kappa\ln 2\right)+\mathcal{O}(\epsilon^2)\bigg]
        \label{I7form}
\end{align}
Here the normalization factor is  
\begin{equation}
    \mathcal{N}=\pi^{-\epsilon} e^{-\gamma_E\epsilon} \left(\frac{2}{3}\right)^{2\epsilon} \frac{1}{q^{1+2\epsilon}}
\end{equation}
with
\begin{equation}
     c_1=\text{Im}\left[\text{Li}_3\left(\frac{i}{\sqrt{3}}\right)\right],
     \quad c_2=\text{Im}\left[\text{Li}_3\left(1+i\sqrt{3}\right)\right],
\end{equation}
and
\begin{align}
      c_3&=-0.949789416853385,\quad
    c_4=-0.305492372030520,\quad c_5=-1.34784474998184,\notag\\ c_6&=4.39528715012114,\quad
    c_7=12.8006547420728,\quad c_8=3.37308464401608
\end{align}

As a cross check, we can add the three sextants in one hemisphere with the same axis projection and compare to the hemisphere soft function~\cite{Fleming:2007xt,Becher:2009th,Chien:2010kc,Kelley:2011ng,Hornig:2011iu}.
\begin{equation}
    I_{\bar{n}_1,\bar{n}_1,n_1}(q)
    =I_1(q)+I_{\bar{n}_1,\bar{n}_3,n_1}(q)+I_{\bar{n}_1,\bar{n}_2,n_1}(q)
\end{equation}
The extra two soft integrals are
\begin{align}
    I_{\bar{n}_1,\bar{n}_3,n_1}(q)&=\mathcal{N}\bigg[{ \frac{3}{2}\ln 2-\frac{3}{2\pi}\kappa}+\epsilon\bigg(\frac{71}{540}\pi^2+\frac{18}{5\pi}c_1-\frac{4}{\pi}c_2+\frac{3}{\pi}\kappa \ln 2-\frac{8}{3}\ln^2 2\notag\\
    &+\frac{3}{2}\ln 2\ln 3-\frac{3}{40}\ln^2 3-\frac{13}{12}\text{Li}_2\left(\frac{1}{4}\right)\bigg)+\mathcal{O}(\epsilon^2)\bigg]\notag\\
    I_{\bar{n}_1,\bar{n}_2,n_1}(q)&=\mathcal{N}\bigg[{ \ln 2+\frac{3}{\pi}\kappa}+\epsilon\bigg(\frac{119}{270}\pi^2-\frac{36}{5\pi}c_1-\frac{4}{\pi}c_2-\frac{6}{\pi}\kappa \ln 2-\frac{2}{3}\ln^2 2\notag\\
    &+\ln 2\ln 3+\frac{3}{20}\ln^2 3+\frac{1}{6}\text{Li}_2\left(\frac{1}{4}\right)\bigg)+\mathcal{O}(\epsilon^2)\bigg]
\end{align}
which leads to the same hemisphere soft function as in Ref.  \cite{Becher:2009th}.

Another interesting fact is that the divergent part of our trijet soft function does not depend on the projection vector $N$. The soft function integral is
\begin{equation}
    I_{n_a,n_b,n_c,n_d,N} (q) \sim \int d^d k  \frac{{n_a} \cdot {n_b}}{({n_a} \cdot
  k) ({n_b} \cdot k)} \delta(k^2)
  \delta \left(q - \frac{2}{3}N \cdot k\right) \times\left[\cdots\right]\\
 \end{equation}
  If we rotate the projection vector $N$ to another direction $N^\prime$, then the $\delta$-function transforms as
\begin{equation}
    \delta\left(q-\frac{2}{3}N^\prime\cdot k\right)
    =\delta\left(q-\frac{2}{3}(N\cdot k)\frac{N^\prime\cdot k}{N\cdot k}\right)
    =\frac{N\cdot k}{N^\prime \cdot k}\delta\left(\frac{N\cdot k}{N^\prime \cdot k} q-\frac{2}{3}N\cdot k\right)
\end{equation}
Then after rescaling 
\begin{equation}
    k \to \frac{N\cdot q}{N^\prime \cdot q}  k
\end{equation}
the integral becomes
\begin{equation}
    I_{n_a,n_b,n_c,n_d,N} (q) \sim \left( \frac{N\cdot q}{N^\prime \cdot q}\right)^{2\epsilon} q^{d-5} \int d^d k  \frac{{n_a} \cdot {n_b}}{({n_a} \cdot
  k) ({n_b} \cdot k)} \delta(k^2)
  \delta \left(q - \frac{2}{3}N \cdot k\right) \times\left[\cdots\right]\\
 \end{equation}
 So the effect of using different $N$'s only shows up at order $\epsilon$ in the expansion. This only affects the anomalous dimension for the integrals which have soft-collinear divergences. This is only $I_1(q)$, since that is the only integral where a Wilson line is in within the integration region. 
 However, for $I_1(q)$ the projection is on $n_1$ (the Wilson line direction) in both thrust and heavy jet mass.  For the others, using the definitions in Eqs.\eqref{I12def}, \eqref{I34def} and \eqref{I567def}, we see that at NLL level after rescaling the projection vector $N$
 \begin{equation}
     I_5(q) = I_3(q),\quad I_6(q) = I_4(q),\quad \text{and} \quad I_7(q) = I_6(q)
 \end{equation}   
 where we need a reflection with respect to $z$-axis to see the third equation. This agrees with our explicit calculations in Eqs.\eqref{I1form}-\eqref{I7form} to order $\epsilon^0$. Note however, that the $\epsilon^1$ terms differ, as expected.
  In summary, at the NLL level, the thrust and heavy jet mass trijet soft function can be taken to be the same. For NNLL resummation and beyond, they will generically be different.

\bibliographystyle{utphys}

\bibliography{sudakov}

\providecommand{\href}[2]{#2}\begingroup\raggedright\begin{thebibliography}{10}

\bibitem{Catani:1997xc}
S.~Catani and B.~R. Webber, ``{Infrared safe but infinite: Soft gluon
  divergences inside the physical region},''
  \href{http://dx.doi.org/10.1088/1126-6708/1997/10/005}{{\em JHEP} {\bfseries
  10} (1997) 005}, \href{http://arxiv.org/abs/hep-ph/9710333}{{\ttfamily
  arXiv:hep-ph/9710333}}.

\bibitem{Seymour:1997kj}
M.~H. Seymour, ``{Jet shapes in hadron collisions: Higher orders, resummation
  and hadronization},''
  \href{http://dx.doi.org/10.1016/S0550-3213(97)00711-6}{{\em Nucl. Phys. B}
  {\bfseries 513} (1998) 269--300},
  \href{http://arxiv.org/abs/hep-ph/9707338}{{\ttfamily arXiv:hep-ph/9707338}}.

\bibitem{Luisoni:2020efy}
G.~Luisoni, P.~F. Monni, and G.~P. Salam, ``{$C$-parameter hadronisation in the
  symmetric 3-jet limit and impact on $\alpha_s$ fits},''
  \href{http://dx.doi.org/10.1140/epjc/s10052-021-08941-z}{{\em Eur. Phys. J.
  C} {\bfseries 81} no.~2, (2021) 158},
  \href{http://arxiv.org/abs/2012.00622}{{\ttfamily arXiv:2012.00622
  [hep-ph]}}.

\bibitem{Benkendorfer:2021unv}
K.~Benkendorfer and A.~J. Larkoski, ``{Grooming at the cusp: all-orders
  predictions for the transition region of jet groomers},''
  \href{http://dx.doi.org/10.1007/JHEP11(2021)188}{{\em JHEP} {\bfseries 11}
  (2021) 188}, \href{http://arxiv.org/abs/2108.02779}{{\ttfamily
  arXiv:2108.02779 [hep-ph]}}.

\bibitem{Farhi:1977sg}
E.~Farhi, ``{A QCD Test for Jets},''
  \href{http://dx.doi.org/10.1103/PhysRevLett.39.1587}{{\em Phys. Rev. Lett.}
  {\bfseries 39} (1977) 1587--1588}.

\bibitem{Salam:2001bd}
G.~P. Salam and D.~Wicke, ``{Hadron masses and power corrections to event
  shapes},'' \href{http://dx.doi.org/10.1088/1126-6708/2001/05/061}{{\em JHEP}
  {\bfseries 05} (2001) 061},
  \href{http://arxiv.org/abs/hep-ph/0102343}{{\ttfamily arXiv:hep-ph/0102343}}.

\bibitem{Chien:2010kc}
Y.-T. Chien and M.~D. Schwartz, ``{Resummation of heavy jet mass and comparison
  to LEP data},'' \href{http://dx.doi.org/10.1007/JHEP08(2010)058}{{\em JHEP}
  {\bfseries 08} (2010) 058}, \href{http://arxiv.org/abs/1005.1644}{{\ttfamily
  arXiv:1005.1644 [hep-ph]}}.

\bibitem{Gehrmann-DeRidder:2007vsv}
A.~Gehrmann-De~Ridder, T.~Gehrmann, E.~W.~N. Glover, and G.~Heinrich, ``{NNLO
  corrections to event shapes in e+ e- annihilation},''
  \href{http://dx.doi.org/10.1088/1126-6708/2007/12/094}{{\em JHEP} {\bfseries
  12} (2007) 094}, \href{http://arxiv.org/abs/0711.4711}{{\ttfamily
  arXiv:0711.4711 [hep-ph]}}.

\bibitem{Catani:1996jh}
S.~Catani and M.~H. Seymour, ``{The Dipole formalism for the calculation of QCD
  jet cross-sections at next-to-leading order},''
  \href{http://dx.doi.org/10.1016/0370-2693(96)00425-X}{{\em Phys. Lett. B}
  {\bfseries 378} (1996) 287--301},
  \href{http://arxiv.org/abs/hep-ph/9602277}{{\ttfamily arXiv:hep-ph/9602277}}.

\bibitem{Catani:1996vz}
S.~Catani and M.~H. Seymour, ``{A General algorithm for calculating jet
  cross-sections in NLO QCD},''
  \href{http://dx.doi.org/10.1016/S0550-3213(96)00589-5}{{\em Nucl. Phys. B}
  {\bfseries 485} (1997) 291--419},
  \href{http://arxiv.org/abs/hep-ph/9605323}{{\ttfamily arXiv:hep-ph/9605323}}.
  [Erratum: Nucl.Phys.B 510, 503--504 (1998)].

\bibitem{Frixione:1998dw}
S.~Frixione, P.~Nason, and G.~Ridolfi, ``{Problems in the resummation of soft
  gluon effects in the transverse momentum distributions of massive vector
  bosons in hadronic collisions},''
  \href{http://dx.doi.org/10.1016/S0550-3213(98)00853-0}{{\em Nucl. Phys. B}
  {\bfseries 542} (1999) 311--328},
  \href{http://arxiv.org/abs/hep-ph/9809367}{{\ttfamily arXiv:hep-ph/9809367}}.

\bibitem{Becher:2010tm}
T.~Becher and M.~Neubert, ``{Drell-Yan Production at Small $q_T$, Transverse
  Parton Distributions and the Collinear Anomaly},''
  \href{http://dx.doi.org/10.1140/epjc/s10052-011-1665-7}{{\em Eur. Phys. J. C}
  {\bfseries 71} (2011) 1665}, \href{http://arxiv.org/abs/1007.4005}{{\ttfamily
  arXiv:1007.4005 [hep-ph]}}.

\bibitem{Monni:2016ktx}
P.~F. Monni, E.~Re, and P.~Torrielli, ``{Higgs Transverse-Momentum Resummation
  in Direct Space},''
  \href{http://dx.doi.org/10.1103/PhysRevLett.116.242001}{{\em Phys. Rev.
  Lett.} {\bfseries 116} no.~24, (2016) 242001},
  \href{http://arxiv.org/abs/1604.02191}{{\ttfamily arXiv:1604.02191
  [hep-ph]}}.

\bibitem{NOGUEIRA1993279}
P.~Nogueira, ``Automatic feynman graph generation,''
  \href{http://dx.doi.org/https://doi.org/10.1006/jcph.1993.1074}{{\em Journal
  of Computational Physics} {\bfseries 105} no.~2, (1993) 279--289}.
  \url{https://www.sciencedirect.com/science/article/pii/S0021999183710740}.

\bibitem{Vermaseren:2000nd}
J.~A.~M. Vermaseren, ``{New features of FORM},''
  \href{http://arxiv.org/abs/math-ph/0010025}{{\ttfamily
  arXiv:math-ph/0010025}}.

\bibitem{HAHN2001418}
T.~Hahn, ``Generating feynman diagrams and amplitudes with feynarts 3,''
  \href{http://dx.doi.org/https://doi.org/10.1016/S0010-4655(01)00290-9}{{\em
  Computer Physics Communications} {\bfseries 140} no.~3, (2001) 418--431}.
  \url{https://www.sciencedirect.com/science/article/pii/S0010465501002909}.

\bibitem{MERTIG1991345}
R.~Mertig, M.~Böhm, and A.~Denner, ``Feyn calc - computer-algebraic
  calculation of feynman amplitudes,''
  \href{http://dx.doi.org/https://doi.org/10.1016/0010-4655(91)90130-D}{{\em
  Computer Physics Communications} {\bfseries 64} no.~3, (1991) 345--359}.
  \url{https://www.sciencedirect.com/science/article/pii/001046559190130D}.

\bibitem{SHTABOVENKO2016432}
V.~Shtabovenko, R.~Mertig, and F.~Orellana, ``New developments in feyncalc
  9.0,''
  \href{http://dx.doi.org/https://doi.org/10.1016/j.cpc.2016.06.008}{{\em
  Computer Physics Communications} {\bfseries 207} (2016) 432--444}.
  \url{https://www.sciencedirect.com/science/article/pii/S0010465516301709}.

\bibitem{Ellis:1996mzs}
R.~K. Ellis, W.~J. Stirling, and B.~R. Webber,
  \href{http://dx.doi.org/10.1017/CBO9780511628788}{{\em {QCD and collider
  physics}}}, vol.~8.
\newblock Cambridge University Press, 2, 2011.

\bibitem{Becher:2009th}
T.~Becher and M.~D. Schwartz, ``{Direct photon production with effective field
  theory},'' \href{http://dx.doi.org/10.1007/JHEP02(2010)040}{{\em JHEP}
  {\bfseries 02} (2010) 040}, \href{http://arxiv.org/abs/0911.0681}{{\ttfamily
  arXiv:0911.0681 [hep-ph]}}.

\bibitem{Schwartz:2016olw}
M.~D. Schwartz, ``{Precision direct photon spectra at high energy and
  comparison to the 8 TeV ATLAS data},''
  \href{http://dx.doi.org/10.1007/JHEP09(2016)005}{{\em JHEP} {\bfseries 09}
  (2016) 005}, \href{http://arxiv.org/abs/1606.02313}{{\ttfamily
  arXiv:1606.02313 [hep-ph]}}.

\bibitem{Becher:2011fc}
T.~Becher, C.~Lorentzen, and M.~D. Schwartz, ``{Resummation for W and Z
  production at large $p_T$},''
  \href{http://dx.doi.org/10.1103/PhysRevLett.108.012001}{{\em Phys. Rev.
  Lett.} {\bfseries 108} (2012) 012001},
  \href{http://arxiv.org/abs/1106.4310}{{\ttfamily arXiv:1106.4310 [hep-ph]}}.

\bibitem{Becher:2012xr}
T.~Becher, C.~Lorentzen, and M.~D. Schwartz, ``{Precision Direct Photon and
  $W$-Boson Spectra at High $p_T$ and Comparison to LHC Data},''
  \href{http://dx.doi.org/10.1103/PhysRevD.86.054026}{{\em Phys. Rev. D}
  {\bfseries 86} (2012) 054026},
  \href{http://arxiv.org/abs/1206.6115}{{\ttfamily arXiv:1206.6115 [hep-ph]}}.

\bibitem{Banfi:2002hw}
A.~Banfi, G.~Marchesini, and G.~Smye, ``{Away from jet energy flow},''
  \href{http://dx.doi.org/10.1088/1126-6708/2002/08/006}{{\em JHEP} {\bfseries
  08} (2002) 006}, \href{http://arxiv.org/abs/hep-ph/0206076}{{\ttfamily
  arXiv:hep-ph/0206076}}.

\bibitem{Schwartz:2014wha}
M.~D. Schwartz and H.~X. Zhu, ``{Nonglobal logarithms at three loops, four
  loops, five loops, and beyond},''
  \href{http://dx.doi.org/10.1103/PhysRevD.90.065004}{{\em Phys. Rev. D}
  {\bfseries 90} no.~6, (2014) 065004},
  \href{http://arxiv.org/abs/1403.4949}{{\ttfamily arXiv:1403.4949 [hep-ph]}}.

\bibitem{Becher:2006mr}
T.~Becher, M.~Neubert, and B.~D. Pecjak, ``{Factorization and Momentum-Space
  Resummation in Deep-Inelastic Scattering},''
  \href{http://dx.doi.org/10.1088/1126-6708/2007/01/076}{{\em JHEP} {\bfseries
  01} (2007) 076}, \href{http://arxiv.org/abs/hep-ph/0607228}{{\ttfamily
  arXiv:hep-ph/0607228}}.

\bibitem{Schwartz:2007ib}
M.~D. Schwartz, ``{Resummation and NLO matching of event shapes with effective
  field theory},'' \href{http://dx.doi.org/10.1103/PhysRevD.77.014026}{{\em
  Phys. Rev. D} {\bfseries 77} (2008) 014026},
  \href{http://arxiv.org/abs/0709.2709}{{\ttfamily arXiv:0709.2709 [hep-ph]}}.

\bibitem{Becher:2008cf}
T.~Becher and M.~D. Schwartz, ``{A precise determination of $\alpha_s$ from LEP
  thrust data using effective field theory},''
  \href{http://dx.doi.org/10.1088/1126-6708/2008/07/034}{{\em JHEP} {\bfseries
  07} (2008) 034}, \href{http://arxiv.org/abs/0803.0342}{{\ttfamily
  arXiv:0803.0342 [hep-ph]}}.

\bibitem{Fleming:2007xt}
S.~Fleming, A.~H. Hoang, S.~Mantry, and I.~W. Stewart, ``{Top Jets in the Peak
  Region: Factorization Analysis with NLL Resummation},''
  \href{http://dx.doi.org/10.1103/PhysRevD.77.114003}{{\em Phys. Rev. D}
  {\bfseries 77} (2008) 114003},
  \href{http://arxiv.org/abs/0711.2079}{{\ttfamily arXiv:0711.2079 [hep-ph]}}.

\bibitem{Becher:2007ty}
T.~Becher, M.~Neubert, and G.~Xu, ``{Dynamical Threshold Enhancement and
  Resummation in Drell-Yan Production},''
  \href{http://dx.doi.org/10.1088/1126-6708/2008/07/030}{{\em JHEP} {\bfseries
  07} (2008) 030}, \href{http://arxiv.org/abs/0710.0680}{{\ttfamily
  arXiv:0710.0680 [hep-ph]}}.

\bibitem{Stewart:2010tn}
I.~W. Stewart, F.~J. Tackmann, and W.~J. Waalewijn, ``{N-Jettiness: An
  Inclusive Event Shape to Veto Jets},''
  \href{http://dx.doi.org/10.1103/PhysRevLett.105.092002}{{\em Phys. Rev.
  Lett.} {\bfseries 105} (2010) 092002},
  \href{http://arxiv.org/abs/1004.2489}{{\ttfamily arXiv:1004.2489 [hep-ph]}}.

\bibitem{Jouttenus:2011wh}
T.~T. Jouttenus, I.~W. Stewart, F.~J. Tackmann, and W.~J. Waalewijn, ``{The
  Soft Function for Exclusive N-Jet Production at Hadron Colliders},''
  \href{http://dx.doi.org/10.1103/PhysRevD.83.114030}{{\em Phys. Rev. D}
  {\bfseries 83} (2011) 114030},
  \href{http://arxiv.org/abs/1102.4344}{{\ttfamily arXiv:1102.4344 [hep-ph]}}.

\bibitem{Feige:2012vc}
I.~Feige, M.~D. Schwartz, I.~W. Stewart, and J.~Thaler, ``{Precision Jet
  Substructure from Boosted Event Shapes},''
  \href{http://dx.doi.org/10.1103/PhysRevLett.109.092001}{{\em Phys. Rev.
  Lett.} {\bfseries 109} (2012) 092001},
  \href{http://arxiv.org/abs/1204.3898}{{\ttfamily arXiv:1204.3898 [hep-ph]}}.

\bibitem{Ellis:1980wv}
R.~K. Ellis, D.~A. Ross, and A.~E. Terrano, ``{The Perturbative Calculation of
  Jet Structure in e+ e- Annihilation},''
  \href{http://dx.doi.org/10.1016/0550-3213(81)90165-6}{{\em Nucl. Phys. B}
  {\bfseries 178} (1981) 421--456}.

\bibitem{Becher:2009qa}
T.~Becher and M.~Neubert, ``{On the Structure of Infrared Singularities of
  Gauge-Theory Amplitudes},''
  \href{http://dx.doi.org/10.1088/1126-6708/2009/06/081}{{\em JHEP} {\bfseries
  06} (2009) 081}, \href{http://arxiv.org/abs/0903.1126}{{\ttfamily
  arXiv:0903.1126 [hep-ph]}}. [Erratum: JHEP 11, 024 (2013)].

\bibitem{Hoang:2008fs}
A.~H. Hoang and S.~Kluth, ``{Hemisphere Soft Function at O($\alpha_s^2$) for
  Dijet Production in $e^+ e^-$ Annihilation},''
  \href{http://arxiv.org/abs/0806.3852}{{\ttfamily arXiv:0806.3852 [hep-ph]}}.

\bibitem{Dasgupta:2001sh}
M.~Dasgupta and G.~P. Salam, ``{Resummation of nonglobal QCD observables},''
  \href{http://dx.doi.org/10.1016/S0370-2693(01)00725-0}{{\em Phys. Lett. B}
  {\bfseries 512} (2001) 323--330},
  \href{http://arxiv.org/abs/hep-ph/0104277}{{\ttfamily arXiv:hep-ph/0104277}}.

\bibitem{Becher:2015hka}
T.~Becher, M.~Neubert, L.~Rothen, and D.~Y. Shao, ``{Effective Field Theory for
  Jet Processes},''
  \href{http://dx.doi.org/10.1103/PhysRevLett.116.192001}{{\em Phys. Rev.
  Lett.} {\bfseries 116} no.~19, (2016) 192001},
  \href{http://arxiv.org/abs/1508.06645}{{\ttfamily arXiv:1508.06645
  [hep-ph]}}.

\bibitem{Larkoski:2015zka}
A.~J. Larkoski, I.~Moult, and D.~Neill, ``{Non-Global Logarithms,
  Factorization, and the Soft Substructure of Jets},''
  \href{http://dx.doi.org/10.1007/JHEP09(2015)143}{{\em JHEP} {\bfseries 09}
  (2015) 143}, \href{http://arxiv.org/abs/1501.04596}{{\ttfamily
  arXiv:1501.04596 [hep-ph]}}.

\bibitem{Abbate:2010xh}
R.~Abbate, M.~Fickinger, A.~H. Hoang, V.~Mateu, and I.~W. Stewart, ``{Thrust at
  $N^{3}LL$ with Power Corrections and a Precision Global Fit for
  $\alpha_{s}(mZ)$},'' \href{http://dx.doi.org/10.1103/PhysRevD.83.074021}{{\em
  Phys. Rev. D} {\bfseries 83} (2011) 074021},
  \href{http://arxiv.org/abs/1006.3080}{{\ttfamily arXiv:1006.3080 [hep-ph]}}.

\bibitem{Catani:1996yz}
S.~Catani, M.~L. Mangano, P.~Nason, and L.~Trentadue, ``{The Resummation of
  soft gluons in hadronic collisions},''
  \href{http://dx.doi.org/10.1016/0550-3213(96)00399-9}{{\em Nucl. Phys. B}
  {\bfseries 478} (1996) 273--310},
  \href{http://arxiv.org/abs/hep-ph/9604351}{{\ttfamily arXiv:hep-ph/9604351}}.

\bibitem{Banfi:2018mcq}
A.~Banfi, B.~K. El-Menoufi, and P.~F. Monni, ``{The Sudakov radiator for jet
  observables and the soft physical coupling},''
  \href{http://dx.doi.org/10.1007/JHEP01(2019)083}{{\em JHEP} {\bfseries 01}
  (2019) 083}, \href{http://arxiv.org/abs/1807.11487}{{\ttfamily
  arXiv:1807.11487 [hep-ph]}}.

\bibitem{Cal:2019hjc}
P.~Cal, F.~Ringer, and W.~J. Waalewijn, ``{The jet shape at NLL'},''
  \href{http://dx.doi.org/10.1007/JHEP05(2019)143}{{\em JHEP} {\bfseries 05}
  (2019) 143}, \href{http://arxiv.org/abs/1901.06389}{{\ttfamily
  arXiv:1901.06389 [hep-ph]}}.

\bibitem{Catani:1992ua}
S.~Catani, L.~Trentadue, G.~Turnock, and B.~R. Webber, ``{Resummation of large
  logarithms in e+ e- event shape distributions},''
  \href{http://dx.doi.org/10.1016/0550-3213(93)90271-P}{{\em Nucl. Phys. B}
  {\bfseries 407} (1993) 3--42}.

\bibitem{Becher:2011xn}
T.~Becher, M.~Neubert, and D.~Wilhelm, ``{Electroweak Gauge-Boson Production at
  Small $q_T$: Infrared Safety from the Collinear Anomaly},''
  \href{http://dx.doi.org/10.1007/JHEP02(2012)124}{{\em JHEP} {\bfseries 02}
  (2012) 124}, \href{http://arxiv.org/abs/1109.6027}{{\ttfamily arXiv:1109.6027
  [hep-ph]}}.

\bibitem{Ebert:2016gcn}
M.~A. Ebert and F.~J. Tackmann, ``{Resummation of Transverse Momentum
  Distributions in Distribution Space},''
  \href{http://dx.doi.org/10.1007/JHEP02(2017)110}{{\em JHEP} {\bfseries 02}
  (2017) 110}, \href{http://arxiv.org/abs/1611.08610}{{\ttfamily
  arXiv:1611.08610 [hep-ph]}}.

\bibitem{Chen:2020vvp}
H.~Chen, I.~Moult, X.~Zhang, and H.~X. Zhu, ``{Rethinking jets with energy
  correlators: Tracks, resummation, and analytic continuation},''
  \href{http://dx.doi.org/10.1103/PhysRevD.102.054012}{{\em Phys. Rev. D}
  {\bfseries 102} no.~5, (2020) 054012},
  \href{http://arxiv.org/abs/2004.11381}{{\ttfamily arXiv:2004.11381
  [hep-ph]}}.

\bibitem{Kelley:2011ng}
R.~Kelley, M.~D. Schwartz, R.~M. Schabinger, and H.~X. Zhu, ``{The two-loop
  hemisphere soft function},''
  \href{http://dx.doi.org/10.1103/PhysRevD.84.045022}{{\em Phys. Rev. D}
  {\bfseries 84} (2011) 045022},
  \href{http://arxiv.org/abs/1105.3676}{{\ttfamily arXiv:1105.3676 [hep-ph]}}.

\bibitem{Hornig:2011iu}
A.~Hornig, C.~Lee, I.~W. Stewart, J.~R. Walsh, and S.~Zuberi, ``{Non-global
  Structure of the $\mathcal{O}({\alpha}^2_s)$ Dijet Soft Function},''
  \href{http://dx.doi.org/10.1007/JHEP08(2011)054}{{\em JHEP} {\bfseries 08}
  (2011) 054}, \href{http://arxiv.org/abs/1105.4628}{{\ttfamily arXiv:1105.4628
  [hep-ph]}}. [Erratum: JHEP 10, 101 (2017)].

\end{thebibliography}\endgroup


\providecommand{\href}[2]{#2}\begingroup\raggedright\endgroup

\end{document}